\pdfoutput=1
\documentclass[pdflatex,sn-mathphys-num]{sn-jnl}

\usepackage{graphicx}
\usepackage{multirow}
\usepackage{amsmath,amssymb,amsfonts}
\usepackage{amsthm}
\usepackage{mathrsfs}
\usepackage[title]{appendix}
\usepackage{xcolor}
\usepackage{textcomp}
\usepackage{manyfoot}
\usepackage{booktabs}
\usepackage{array}
\usepackage{tikz}
\usetikzlibrary{arrows.meta,positioning,shapes.geometric}
\graphicspath{{image/}}

\begin{document}

\title[Choice-Implied Subjective Time]{Functional Characterization of Choice-Implied Subjective Time in Intertemporal Financial Decisions}

\author*[1]{\fnm{Fabrizio} \sur{Maturo}}\email{fabrizio.maturo@unimercatorum.it}

\author[2]{\fnm{Salvador} \sur{Cruz Rambaud}}\email{scruz@ual.es}

\author[3]{\fnm{Vincenzo} \sur{Li Calzi}}\email{vincenzo.licalzi@studenti.unimercatorum.it}

\author[1]{\fnm{Andrea} \sur{Mazzitelli}}\email{a.mazzitelli@unimercatorum.it}

\author[4,5]{\fnm{Annamaria} \sur{Porreca}}\email{annamaria.porreca@uniroma5.it}

\affil*[1]{\orgdiv{Department of Economics, Statistics and Business, Faculty of Technological and Innovation Sciences}, \orgname{Universitas Mercatorum}, \orgaddress{\city{Rome}, \country{Italy}}}

\affil[2]{\orgdiv{Departamento de Econom\'{i}a y Empresa}, \orgname{Universidad de Almer\'{i}a}, \orgaddress{\city{Almer\'{i}a}, \country{Spain}}}

\affil[3]{\orgdiv{Department of Engineering and Science}, \orgname{Universitas Mercatorum}, \orgaddress{\city{Rome}, \country{Italy}}}

\affil[4]{\orgdiv{Department for the Promotion of Human Science and Quality of Life}, \orgname{San Raffaele University}, \orgaddress{\city{Rome}, \country{Italy}}}

\affil[5]{\orgdiv{Unit of Clinical and Molecular Epidemiology}, \orgname{IRCCS San Raffaele Roma}, \orgaddress{\city{Rome}, \country{Italy}}}

\abstract{Intertemporal choice is usually summarized by a discount rate or a predefined parametric discount function, leaving differences in the shape of individual discounting over the horizon unobserved. This paper proposes a Functional Data Analysis framework that reconstructs, for each respondent, a choice-implied subjective-time trajectory $\tau_i(t)$ from monetary equivalence judgments collected at thirteen horizons between seven days and five years. Equivalence judgments are mapped into discount factors, regularized by monotone shape-constrained smoothing, and converted into trajectories under the representation $f_i(t)=\exp\{-k_i\tau_i(t)\}$, with $k_i$ an operational scale. The trajectories are analyzed through derivatives, functional principal component analysis and clustering of the component scores, compared with exponential, Weber--Fechner and Stevens benchmarks in sample and in a leave-one-horizon-out design, and related to an explicit subjective-time positioning task. A Monte Carlo study evaluates the recovery of trajectory shape and group structure. The empirical application uses an online multilingual survey with 141 respondents; 140 are eligible, 61 report the full future amount at every horizon and 19 provide profiles that do not determine a decreasing discount curve, leaving 60 identified trajectories. The mean trajectory is predominantly concave, two functional components account for 92.0\% of the variability of the centered trajectories, and four profiles that differ in when subjective time accumulates relative to the mean are identified; the profiles also differ in discounting intensity. Parametric subjective-time models are competitive in predicting held-out judgments, while partitions based on their parameters do not reproduce the functional partition. No statistically detectable correspondence with the explicit positioning task is found.}


\maketitle

\section{Introduction}
\label{sec:intro}

Intertemporal choices require individuals to compare outcomes that differ both in magnitude and in the time at which they are received \citep{Loewenstein_1989}. The discounted utility model represents this comparison through an exponential discount function defined on chronological time, so that preferences are summarized by a constant discount rate \citep{Samuelson_1937}. A large body of evidence has documented systematic departures from this benchmark, including decreasing impatience, preference reversals and intransitive choice patterns \citep{Thaler_1981,Loewenstein_1992,Roelofsma_2000,Cruz_Rambaud_2019}. One line of explanation attributes these departures to the way individuals represent temporal distance: if the exponential structure is retained but applied to a subjective transformation of time rather than to calendar time, non-exponential discounting in chronological time can arise from a nonlinear perception of delay \citep{Zauberman_2009,Kim_2009,Cruz_Rambaud_2023}. The object of interest then becomes the transformation of chronological into subjective time instead of a single rate parameter.

Empirical work in this direction has mostly relied on parametric transformations, typically logarithmic (Weber--Fechner) or power-law (Stevens) specifications \citep{Takahashi_2008,Zauberman_2009}. These forms are parsimonious and interpretable, but they impose a common functional family on all individuals. When equivalence judgments are observed at many horizons, the information contained in the data concerns the whole trajectory: where along the horizon impatience is concentrated, whether subjective time accumulates early or late, and whether the transformation changes curvature. Functional Data Analysis (FDA) treats each individual sequence of observations as the discrete realization of a smooth function and is therefore a natural tool for this type of data \citep{Ramsay_2005}. FDA has been applied to discount functions and their derivatives \citep{Martino_2024,Ventre_2024} and to the clustering of discount functions \citep{Maturo_2026}. In these works the functional object is the discount function itself, so that the intensity of discounting and the shape of its time profile enter the analysis jointly.

The present paper targets a different functional object. Starting from monetary equivalence judgments, we reconstruct for each respondent a choice-implied subjective-time trajectory $\tau_i(t)$ under the representation $f_i(t)=\exp\{-k_i\tau_i(t)\}$, where $k_i$ is fixed by an explicit operational rule. Since only the product $k_i\tau_i(t)$ is determined by choice data, $\tau_i(t)$ is a normalized trajectory whose shape, rather than its absolute level, is the quantity of interest. The representation is a transformation of the discount function that separates an operational scale of discounting intensity from the shape of the deviations from exponential discounting, and the latter are analyzed as functional data. The reconstruction combines two monotone shape-constrained smoothing steps with an explicit treatment of the profiles for which the trajectory is not determined. The reconstructed trajectories are described through derivatives, functional principal component analysis (FPCA) and clustering of the component scores, and the resulting profiles are related to discounting intensity and to individual covariates. The functional reconstruction is compared with exponential, Weber--Fechner and Stevens benchmarks in sample and in a leave-one-horizon-out design, and the recoverability of the functional profiles from the parameters of the parametric models is examined. The implicit trajectories are then compared with an explicit subjective-time positioning task through permutation tests. A Monte Carlo study with known generating trajectories evaluates how well the procedure recovers trajectory shape, low-dimensional structure and group membership.

Section~\ref{sec:background} summarizes the theoretical link between subjective time and discounting. Section~\ref{sec:data} describes the questionnaire and the construction of the analytical sample. Section~\ref{sec:method} presents the functional framework, Section~\ref{sec:simulation} the simulation study and Section~\ref{sec:results} the empirical results. Section~\ref{sec:discussion} discusses the findings and concludes.

\section{Subjective time and discounting}
\label{sec:background}

Let $g(t)$ be a time distortion, that is a continuous and strictly increasing function on $[0,t_0)$ with $g(0)=0$, which maps chronological time into subjective time \citep{Cruz_Rambaud_2023}. Retaining the exponential structure on the subjective scale gives
\begin{equation}
\label{eq:Exponential}
F(t)=\exp\{-k\,g(t)\},\qquad k>0,
\end{equation}
which reduces to the classical exponential model when $g(t)=t$. When $g$ is nonlinear, discounting is exponential in subjective time but generally non-exponential in chronological time \citep{Cruz_Rambaud_2023,Cruz_Rambaud_2023_Review}. The instantaneous discount rate is
\begin{equation}
\delta(t)=-\frac{\mathrm{d}}{\mathrm{d}t}\ln F(t)=k\,g'(t),
\end{equation}
so that the evolution of impatience is governed by the speed $g'(t)$ at which subjective time accumulates. Decreasing impatience corresponds to a log-convex discount function, which under \eqref{eq:Exponential} is equivalent to a concave $g$ \citep{Prelec_2004}; the local variation of impatience can be summarized by Prelec's index $P(t)=-\mathrm{d}\ln\delta(t)/\mathrm{d}t$ \citep{Prelec_2004,Cruz_Rambaud_2020}. The derivative $g'(t)$ coincides with the rate of subjective time introduced in the psychological literature \citep{McGrath_1967}. Equivalent representations of the same preferences can be given through delay functions \citep{Cruz_Rambaud_2020}, and the separate roles of interval length and interval location in the perception of delay have been documented \citep{Kim_2019}.

Two parametric families dominate the literature. The Weber--Fechner transformation
\begin{equation}
\label{eq:weber}
g_1(t)=\alpha\ln(1+\beta t),\qquad \alpha,\beta>0,
\end{equation}
is increasing and concave and, inserted in \eqref{eq:Exponential}, yields the generalized hyperbolic discount function $F(t)=(1+\beta t)^{-k\alpha}$, which reduces to the hyperbolic form when $k\alpha=1$ \citep{Takahashi_2008,Cruz_Rambaud_2023}. The Stevens transformation
\begin{equation}
\label{eq:Stevens}
g_2(t)=\alpha t^{\beta},\qquad \alpha,\beta>0,
\end{equation}
is concave for $\beta<1$, linear for $\beta=1$ and convex for $\beta>1$ \citep{Takahashi_2008}. Models of subjective temporal compression in which the subjective unit decreases with elapsed time lead to logarithmic or power-law forms depending on a single compression parameter \citep{Galam_2024}. Direct measurements, in which respondents position future horizons on a line from the present to the far future, show that perceived duration grows less than proportionally with calendar time and that discounting computed on perceived time is closer to exponential \citep{Zauberman_2009}. Reviews of this literature emphasize that the mapping from chronological to perceived time varies across contexts and individuals \citep{Zauberman_2011,Zauberman_2016}, and experimental work shows that discounting and explicit time perception can be affected in dissociable ways \citep{Xu_2020}. These results motivate reconstructing the individual transformation from the data, with the parametric families used as benchmarks.

\section{Data}
\label{sec:data}

\subsection{Questionnaire}

Data were collected through a multilingual online questionnaire, available in Italian, English and Spanish, presented as a university study on how individuals perceive time and evaluate future payments. Respondents were recruited between February and September 2026 through a non-probability convenience procedure based on personal contacts, contacts of initial recipients and dissemination of the link on social media. Participation was voluntary and respondents were told that there were no correct or incorrect answers.

The implicit component consists of monetary equivalence judgments. For each horizon, respondents indicated the amount received immediately that they considered equivalent to receiving $100$ euros at that horizon. Thirteen horizons were used: $7$, $14$ and $21$ days; $1$, $2$, $3$, $4$, $6$ and $9$ months; and $1$, $2$, $3$ and $5$ years. The explicit component is a subjective time positioning (STP) task in which respondents placed each of the same thirteen horizons on a line from the present to the far future, following \citet{Zauberman_2009}. The questionnaire also collected a Temporal Sense Scale (TSS), in which respondents rated the perceived speed of time from $1$ (very slowly) to $7$ (very quickly) for several time windows, three financial literacy items on interest compounding, inflation and diversification, and socio-demographic information. Categorical responses from the three language versions were harmonized into common categories, the TSS items were recoded on the $1$--$7$ scale, and the core TSS index (TSS$_{\text{core}}$) was computed as the mean of the items referring to a day, a week, a month, a year and one's life as a whole. The financial literacy score is the number of correct answers ($0$--$3$).

\subsection{Analytical sample}

The questionnaire was completed by $141$ respondents. One respondent provided only four valid equivalence judgments and was excluded, because individual smoothing requires at least six valid observations; the remaining $140$ respondents form the eligible sample described in Table~\ref{tab:sample}. Values $f_i(t)\le 0$ or $f_i(t)>1$ were treated as missing; no equivalence judgment or STP value was missing in the eligible sample.

\begin{table}[htbp]
\centering
\caption{Characteristics of the eligible sample ($N=140$). Categories are harmonized across the three language versions of the questionnaire.}
\label{tab:sample}
\footnotesize
\begin{tabular}{llcc}
\toprule
Variable & Category & $n$ & \% \\
\midrule
\multirow{2}{*}{Gender} & Female & 78 & 55.7 \\
 & Male & 62 & 44.3 \\
\midrule
\multirow{6}{*}{Education} & Lower secondary & 3 & 2.1 \\
 & Upper secondary & 42 & 30.0 \\
 & Bachelor & 23 & 16.4 \\
 & Master's degree & 57 & 40.7 \\
 & PhD & 12 & 8.6 \\
 & Other & 3 & 2.1 \\
\midrule
\multirow{6}{*}{Employment} & Full-time & 59 & 42.1 \\
 & Part-time & 27 & 19.3 \\
 & Student & 19 & 13.6 \\
 & Self-employed & 15 & 10.7 \\
 & Other & 16 & 11.4 \\
 & Retired & 4 & 2.9 \\
\midrule
\multirow{6}{*}{Income (EUR/year)} & $<10{,}000$ & 27 & 19.3 \\
 & $10{,}000$--$20{,}000$ & 27 & 19.3 \\
 & $20{,}001$--$40{,}000$ & 40 & 28.6 \\
 & $40{,}001$--$60{,}000$ & 12 & 8.6 \\
 & $60{,}001$--$100{,}000$ & 5 & 3.6 \\
 & Prefer not to say & 29 & 20.7 \\
\midrule
\multicolumn{4}{l}{Age (years): mean $39.5$, SD $12.0$, range $18$--$73$}\\
\multicolumn{4}{l}{TSS$_{\text{core}}$: mean $4.85$, SD $1.15$}\\
\multicolumn{4}{l}{Financial literacy ($0$--$3$): mean $1.80$, SD $0.90$}\\
\bottomrule
\end{tabular}
\end{table}

Among the $140$ eligible respondents, $61$ ($43.6\%$) reported an immediate equivalent of $100$ euros at every horizon, so that $f_i(t)=1$ throughout. For these exact-flat profiles $-\log f_i(t)=0$ at every horizon and the representation $f_i(t)=\exp\{-k_i\tau_i(t)\}$ does not determine $\tau_i(t)$ (Section~\ref{sec:identification}); the classification is based on the raw responses, not on the smoothed curves. Among the remaining $79$ respondents, $19$ provided profiles for which the monotone decreasing fit of Section~\ref{sec:smoothing} is constant and equal to one after normalization, so that the estimated scale is $k_i<10^{-8}$ and $\tau_i(t)$ is again not determined. In $18$ of these $19$ raw profiles the last judgment is equal to or higher than the first, and $18$ show a positive rank correlation between horizon and equivalent amount, that is, the immediate equivalent tends to increase with delay. The functional analysis is therefore based on $N=60$ respondents with identified trajectories. In this sample, $10$ respondents show at least one increase in the equivalent amount between consecutive horizons, which the monotone smoothing step regularizes.

Table~\ref{tab:selection} compares the three groups. Age, financial literacy and TSS$_{\text{core}}$ do not differ significantly across groups, whereas the share of men is higher among respondents with identified trajectories ($63.3\%$) than among exact-flat ($27.9\%$) and non-decreasing ($36.8\%$) respondents. In a logistic model for identification among the $140$ eligible respondents including age, gender, financial literacy and TSS$_{\text{core}}$, the odds ratio for men is $3.68$ ($95\%$ CI $1.71$--$7.89$), while the confidence intervals of the other covariates include one. The empirical results below refer to respondents whose equivalence judgments decrease with delay.

\begin{table}[htbp]
\centering
\caption{Comparison of eligible respondents by reconstruction outcome. Continuous variables: mean (SD) [median], Kruskal--Wallis test. Categorical variables: $n$ (\%), Fisher exact test with simulated $p$-value ($10{,}000$ replicates). Exact-flat: immediate equivalent equal to $100$ at every horizon. Non-decreasing: non-flat profiles whose monotone decreasing fit is constant.}
\label{tab:selection}
\footnotesize
\begin{tabular}{lcccc}
\toprule
 & Identified & Exact-flat & Non-decreasing & $p$ \\
 & ($n=60$) & ($n=61$) & ($n=19$) & \\
\midrule
Age (years) & 41.0 (13.2) [41] & 39.4 (10.8) [38] & 35.1 (11.3) [31] & 0.165 \\
Financial literacy & 1.97 (0.94) [2] & 1.74 (0.83) [2] & 1.47 (0.90) [1] & 0.062 \\
TSS$_{\text{core}}$ & 4.89 (1.13) [5] & 4.86 (1.20) [5] & 4.68 (1.09) [5] & 0.782 \\
Completion time (s), median & 726 & 738 & 632 & 0.706 \\
Men & 38 (63.3) & 17 (27.9) & 7 (36.8) & $<0.001$ \\
PhD & 10 (16.7) & 2 (3.3) & 0 (0.0) & 0.185$^{a}$ \\
Italian version & 55 (91.7) & 58 (95.1) & 18 (94.7) & 0.448$^{a}$ \\
\bottomrule
\end{tabular}
\footnotetext{$^{a}$ Test on the full distribution of education level and of questionnaire language, respectively.}
\end{table}

\section{Functional framework}
\label{sec:method}

Figure~\ref{fig:methodological_pipeline} summarizes the procedure. Each step is applied to every respondent separately up to the construction of $\tau_i(t)$; FPCA and clustering are then applied to the sample of identified trajectories.

\begin{figure}[htbp]
\centering
\resizebox{\textwidth}{!}{%
\begin{tikzpicture}[
  font=\footnotesize,
  node distance=6mm and 6mm,
  box/.style={draw, rounded corners, align=center, minimum width=2.6cm, minimum height=1.05cm, inner sep=2pt, fill=gray!8},
  arr/.style={-{Latex[length=2mm]}, thick}
]
\node[box] (a) {Equivalence\\judgments};
\node[box, right=of a] (b) {Discount factors\\$f_i(t_j)$};
\node[box, right=of b] (c) {Monotone decreasing\\smoothing of $f_i$};
\node[box, right=of c] (d) {Scale $k_i$ and\\trajectory $\tau_i(t)$};
\node[box, below=of d] (e) {Derivatives\\$\tau_i'(t),\ \tau_i''(t)$};
\node[box, left=of e] (f) {FPCA of\\$\tau_i^c(t)$};
\node[box, left=of f] (g) {$k$-means on\\standardized scores};
\node[box, left=of g] (h) {Stability, benchmarks,\\STP comparison};
\draw[arr] (a) -- (b);
\draw[arr] (b) -- (c);
\draw[arr] (c) -- (d);
\draw[arr] (d) -- (e);
\draw[arr] (d.south) -- (f.north);
\draw[arr] (f) -- (g);
\draw[arr] (g) -- (h);
\end{tikzpicture}}
\caption{Steps of the functional framework. Equivalence judgments are converted into discount factors, smoothed under a decreasing constraint and transformed into subjective-time trajectories; the centered trajectories are summarized by FPCA, the standardized component scores are clustered, and the resulting representation is assessed through bootstrap stability, parametric benchmarks and comparison with the explicit STP task.}
\label{fig:methodological_pipeline}
\end{figure}

\subsection{Discount factors and monotone smoothing}
\label{sec:smoothing}

For respondent $i$ and horizon $t_j$, the observed quantity is the immediate amount $A_{0,i}(t_j)$ judged equivalent to $A_f=100$ euros at $t_j$, which defines the discount factor
\begin{equation}
f_i(t_j)=\frac{A_{0,i}(t_j)}{A_f},\qquad j=1,\dots,13.
\end{equation}
The ratio is the discount factor implied by the equivalence judgment under linear utility over the stakes considered, and more generally an operational measure of monetary discounting.
The point $(0,1)$ is added to each respondent's data and the discount curve is estimated by a shape-constrained additive model (SCAM) with a monotone decreasing P-spline basis of dimension $10$ \citep{Pya_2015}. The smoothing parameter is selected by generalized cross-validation (GCV). Because a penalized fit does not interpolate the anchor exactly, the fitted curve is divided by its value at $t=0$, which imposes $\widehat f_i(0)=1$ and preserves monotonicity, and values are restricted to $(0,1]$. The fitted curve $\widehat f_i(t)$ is evaluated on a grid of $200$ equally spaced points on $[0,1825]$ days and at the observed horizons.

\subsection{Subjective-time trajectories, normalization and derivatives}
\label{sec:identification}

Under the representation
\begin{equation}
\label{eq:representation}
f_i(t)=\exp\{-k_i\tau_i(t)\},\qquad\text{equivalently}\qquad -\log f_i(t)=k_i\tau_i(t),
\end{equation}
the discount curve determines only the product $k_i\tau_i(t)$: for any $c>0$, the pair $(ck_i,\tau_i/c)$ produces the same discount factors. A normalization is therefore required. We define $k_i$ as the slope of the regression without intercept of $-\log\widehat f_i(t_j)$ on $t_j$,
\begin{equation}
\label{eq:k}
k_i=\frac{\sum_j t_j\,[-\log\widehat f_i(t_j)]}{\sum_j t_j^2},
\end{equation}
which is the rate of the exponential model in chronological time closest to the smoothed curve in least squares. With this choice, the raw trajectory
\begin{equation}
\label{eq:tauraw}
\tau_{\mathrm{raw},i}(t_j)=\frac{-\log\widehat f_i(t_j)}{k_i}
\end{equation}
equals $t_j$ for an exponential discounter, and departures from the identity describe departures from exponential discounting in chronological time. Substituting \eqref{eq:k} into \eqref{eq:tauraw} gives
\begin{equation}
\label{eq:orthogonality}
\sum_j t_j\,\bigl[\tau_{\mathrm{raw},i}(t_j)-t_j\bigr]=0,
\end{equation}
so that the deviations of each raw trajectory from chronological time are orthogonal to the vector of observed horizons. A trajectory that differs from the identity therefore lies above it over part of the horizon and below it over another part, with longer horizons weighing more in the determination of the scale. The raw trajectory
is then smoothed by a SCAM with a monotone increasing basis of dimension $8$ and GCV smoothing, evaluated on the grid and shifted so that $\tau_i(0)=0$. After this second smoothing, \eqref{eq:representation} and \eqref{eq:orthogonality} hold approximately, and $\tau_i(t)$ is a regularized, non-decreasing reconstruction that may contain flat segments.

Under this normalization, $\tau_i(t)$ is an implicit trajectory conditional on $k_i$, not a direct psychophysical measure of perceived duration, and it includes any component of the equivalence judgments that departs from exponential discounting in chronological time. When $\widehat f_i\equiv 1$, \eqref{eq:k} yields $k_i=0$ and \eqref{eq:tauraw} is undefined. This is the case of the exact-flat profiles and of the non-decreasing profiles of Section~\ref{sec:data}, which are not included in the functional analysis. Since the scale of $\tau_i(t)$ is fixed by the normalization, the information of interest is its shape: where along the horizon subjective time accumulates faster or slower than chronological time and where it crosses the identity.

The first derivative $\tau_i'(t)$ is the local speed of subjective time, the empirical counterpart of $g'(t)$ in Section~\ref{sec:background}; values above one indicate horizons at which subjective time accumulates faster than calendar time. The second derivative $\tau_i''(t)$ describes the change of this speed. Both are computed numerically from the smoothed trajectories evaluated on the grid (spacing of about $9.2$ days) and are used descriptively.

\subsection{Functional principal component analysis}

Let $\bar\tau(t)$ be the sample mean trajectory and $\tau_i^c(t)=\tau_i(t)-\bar\tau(t)$ the centered trajectories. FPCA represents
\begin{equation}
\tau_i^c(t)\approx\sum_{m=1}^{M}\xi_{im}\psi_m(t),
\end{equation}
where $\psi_m$ are orthonormal eigenfunctions of the sample covariance operator and $\xi_{im}$ are the scores \citep{Ramsay_2005}. FPCA is computed on the centered trajectories evaluated on the common grid, with integrals approximated on the grid, using the \texttt{fda.usc} package in R. Before the second smoothing step, \eqref{eq:orthogonality} implies $\sum_j t_j\,\tau_i^c(t_j)=0$ for the centered trajectories, so that the components describe how deviations from the mean are distributed along the horizon under this constraint. The uncertainty of $\bar\tau(t)$ is assessed by a subject-level bootstrap with $B=1000$ resamples, from which $95\%$ pointwise percentile bands and a $95\%$ simultaneous band based on the sup-$t$ statistic are obtained; the origin, where all trajectories equal zero, is excluded from the computation of the sup-$t$ statistic.

\subsection{Clustering of component scores}
\label{sec:clustering}

Groups of respondents with similar trajectory shapes are identified by $k$-means applied to the standardized scores of the first $M$ components, with $M$ the smallest number of components, and at least two, explaining $90\%$ of the variability. Standardization gives each retained component equal weight in the distance; clustering based on the $L^2$ distance between centered trajectories is examined in Section~\ref{sec:sensitivity}. The number of clusters is chosen by the average silhouette over $K=2,\dots,6$, with $50$ random starts. Each cluster is summarized by its functional centroid
\begin{equation}
\bar\tau_r^c(t)=\frac{1}{n_r}\sum_{i\in C_r}\tau_i^c(t),
\end{equation}
whose positive values indicate horizons at which members of cluster $r$ have accumulated more subjective time than the sample mean. Cluster labels are ordered canonically by the value of the centroid at long horizons. Stability is assessed by the mean Jaccard similarity over $500$ bootstrap resamples of respondents, in which $k$-means is re-run on the standardized scores. Clusters are related to $k_i$, which is estimated from the same judgments, and to covariates not used in the clustering (age, gender, education, TSS$_{\text{core}}$, financial literacy, mean STP) through descriptive comparisons.

\subsection{Parametric benchmarks}
\label{sec:benchmark_method}

The functional reconstruction is compared with three parametric discount functions estimated for each respondent by nonlinear least squares on the thirteen observed discount factors: the exponential model $f_i(t)=\exp(-\lambda_it)$, the Weber--Fechner model $f_i(t)=\exp\{-a_i\ln(1+b_it)\}$ and the Stevens model $f_i(t)=\exp(-a_it^{b_i})$, where $a_i$ corresponds to the composite scale $k_i\alpha_i$ of \eqref{eq:weber}--\eqref{eq:Stevens}, the only scale identified by choice data, and $b_i$ to $\beta_i$; for comparison with the functional representation, $\alpha_i=a_i/k_i$ is computed with the $k_i$ of \eqref{eq:k}. Starting values are drawn on a grid with multiple starts. The FDA prediction at horizon $t_j$ is $\exp\{-k_i\tau_i(t_j)\}$. The NLS estimate $\widehat\lambda_i$ differs in general from the scale $k_i$ of \eqref{eq:k}, which is computed on the smoothed curve. Two criteria are used. In-sample accuracy is the root mean squared error (RMSE) and mean absolute error between fitted and observed discount factors at the thirteen horizons. Accuracy on horizons not used in estimation is assessed, for the respondents in the analytical sample, by a leave-one-horizon-out (LOHO) design: for each respondent and each horizon, every model, including the complete FDA reconstruction with re-estimation of $k_i$, is fitted on the remaining twelve horizons and used to predict the held-out discount factor. When the FDA reconstruction without one horizon yields $k_i<10^{-8}$, the smoothed discount curve is used as the prediction. Because the five-year prediction requires extrapolation and the seven-day prediction is informed only by the origin and the following horizons, the LOHO error is also reported for the eleven interior horizons.

Whether the functional partition can be recovered from parametric summaries is assessed by clustering the standardized parameter vectors of the Weber--Fechner and Stevens models, with and without $k_i$, into the same number of clusters, and comparing the resulting partitions with the functional one through the Adjusted Rand Index (ARI).

\subsection{Implicit and explicit subjective time}
\label{sec:stp_method}

The STP task measures stated temporal distance and the trajectory $\tau_i(t)$ is inferred from choices; the two are on different scales and are compared through their shapes. Two permutation tests with $B=5000$ permutations are used. Test T1 asks whether STP profiles differ across the functional clusters: STP values are standardized within respondent, and the statistic is the $F$ statistic for the cluster-by-horizon interaction in a linear model of the standardized values on horizon and cluster; the null distribution is obtained by permuting cluster labels across respondents. A one-way test on the respondent mean STP is also reported. Test T2 asks whether the shape of the cluster centroids of $\tau$ agrees with the shape of the cluster mean STP profiles beyond what is expected when the individual pairing between choices and STP is broken. Both $\tau_i(t)$ at the observed horizons and STP are centered horizon-wise across respondents; for each cluster, $r_k$ is the correlation over horizons between the centroid of centered $\tau$ and the mean centered STP, and the global statistic is $G=\sum_k n_k r_k^2/N$. The null distribution is obtained by permuting complete STP profiles across respondents while keeping trajectories and clusters fixed. Since the seven-day STP value is identical for all respondents, it is zero after centering and the correlations use the remaining twelve horizons. Pearson correlations among $k_i$, the first two FPC scores, TSS$_{\text{core}}$ and mean STP complement the tests.

\subsection{Sensitivity analyses}
\label{sec:sensitivity}

The reference partition is compared through the ARI with partitions obtained by varying: the number of retained components ($2$ to $6$); the clustering algorithm ($k$-means with $25$, $50$ and $100$ starts and Hartigan--Wong, Lloyd and MacQueen algorithms, PAM, Ward.D2); the distance ($L^2$ distance between centered trajectories and semimetrics based on first and second derivatives); the basis dimensions of the two smoothing steps ($8$, $10$, $12$ for $f$ and $6$, $8$, $10$ for $\tau$); the rule for $k_i$, replacing \eqref{eq:k} with the ratio of sums $\tilde k_i=\sum_j[-\log\widehat f_i(t_j)]/\sum_jt_j$; and the exclusion of the three trajectories with the largest $\max_t|\tau_i''(t)|$.

\section{Simulation study}
\label{sec:simulation}

\subsection{Design}

The Monte Carlo study evaluates the reconstruction when the generating trajectory is known. Since $k_i$ and $\tau_i(t)$ are separated by the normalization of Section~\ref{sec:identification}, the study assesses whether the procedure recovers the shape of the trajectories, their low-dimensional structure and their group membership. With $u=t/1825$, four families of monotone trajectories were generated as $\tau(t)=1825\,g(u)$ with
\begin{gather*}
g_{\text{lin}}(u)=u,\qquad g_{\text{conc}}(u)=\frac{\ln(1+12u)}{\ln 13},\qquad g_{\text{conv}}(u)=u^{2.6},\\
g_{\text{sigm}}(u)=\frac{s(u)-s(0)}{s(1)-s(0)},\qquad s(u)=\frac{1}{1+e^{-10(u-0.5)}},
\end{gather*}
so that $\tau(0)=0$ and $\tau(1825)=1825$. Each simulated respondent is assigned to a family with equal probability, $\log k_i$ is drawn from a normal distribution with mean $\log(2\times10^{-4})$ and standard deviation $1.2$, and discount factors $f_i(t_j)=\exp\{-k_i\tau(t_j)\}$ at the thirteen questionnaire horizons are multiplied by $\exp(\varepsilon_{ij})$, with $\varepsilon_{ij}$ normal with mean zero and standard deviation $0.02$, $0.05$ or $0.10$ (low, medium, high noise); values are then truncated to $[10^{-4},1]$. Sample sizes are $N=50,100,140,200$, the value $140$ matching the eligible sample. A stress scenario combines $N=100$, high noise and seven horizons. The reconstruction is identical to the empirical one, including the anchor at $t=0$, both smoothing steps and the rule for $k_i$; in the stress scenario the basis dimensions are adapted to the number of observations. Each scenario was replicated $R=100$ times, with seed $10000+1000s+r$ for scenario $s$ and replication $r$.

Recovery is measured by the RMSE between generating and reconstructed trajectories after dividing each by its own maximum, which makes the comparison invariant to the difference between the endpoint scaling used for generation and the scale $k_i$ used for reconstruction; by the RMSE of the first derivative of the same max-normalized trajectories, computed by finite differences on the grid (per day); by the cumulative variance of the first two FPCs of the generating and of the reconstructed trajectories; and by the ARI between the four generating families and the partition obtained by $k$-means with $K=4$ on the standardized scores of the first two FPCs of the reconstructed trajectories. The number of clusters is fixed at the true value, so the simulation evaluates group recovery and not the selection of $K$.

\subsection{Results}

\begin{table}[htbp]
\centering
\caption{Monte Carlo results: medians over $R=100$ replications, interquartile range in brackets. Shape RMSE: RMSE between max-normalized generating and reconstructed trajectories. Cum.\ var.: variance explained by the first two FPCs of the reconstructed trajectories; for the generating trajectories it is $0.9987$--$0.9988$ in every scenario. ARI: agreement between the four generating families and the four clusters. Stress: $N=100$, high noise, seven horizons.}
\label{tab:simulation_results}
\footnotesize
\setlength{\tabcolsep}{3pt}
\begin{tabular}{@{}lcccc@{}}
\toprule
Scenario & Shape RMSE & RMSE $\tau'(t)$ & Cum.\ var. & ARI \\
\midrule
$N=50$, low & 0.0547 [0.0486, 0.0614] & $2\times10^{-4}$ & 0.9122 & 0.467 [0.410, 0.548] \\
$N=100$, low & 0.0546 [0.0510, 0.0583] & $2\times10^{-4}$ & 0.9082 & 0.491 [0.451, 0.562] \\
$N=140$, low & 0.0566 [0.0525, 0.0609] & $2\times10^{-4}$ & 0.9048 & 0.467 [0.412, 0.528] \\
$N=200$, low & 0.0558 [0.0527, 0.0595] & $2\times10^{-4}$ & 0.9054 & 0.461 [0.417, 0.509] \\
$N=50$, medium & 0.0877 [0.0757, 0.1018] & $3\times10^{-4}$ & 0.9266 & 0.280 [0.214, 0.333] \\
$N=100$, medium & 0.0878 [0.0792, 0.0987] & $3\times10^{-4}$ & 0.9222 & 0.256 [0.217, 0.296] \\
$N=140$, medium & 0.0907 [0.0837, 0.0998] & $3\times10^{-4}$ & 0.9221 & 0.249 [0.199, 0.297] \\
$N=200$, medium & 0.0898 [0.0827, 0.0963] & $3\times10^{-4}$ & 0.9206 & 0.248 [0.217, 0.292] \\
$N=50$, high & 0.1314 [0.1271, 0.1391] & $4\times10^{-4}$ & 0.9362 & 0.132 [0.091, 0.172] \\
$N=100$, high & 0.1312 [0.1290, 0.1359] & $4\times10^{-4}$ & 0.9342 & 0.132 [0.092, 0.167] \\
$N=140$, high & 0.1312 [0.1305, 0.1349] & $4\times10^{-4}$ & 0.9326 & 0.122 [0.096, 0.150] \\
$N=200$, high & 0.1313 [0.1305, 0.1334] & $4\times10^{-4}$ & 0.9332 & 0.131 [0.106, 0.147] \\
Stress & 0.1338 [0.1308, 0.1428] & $4\times10^{-4}$ & 0.9600 & 0.109 [0.077, 0.136] \\
\bottomrule
\end{tabular}
\end{table}

All simulated profiles were reconstructed in every replication, except one profile in one scenario. In Table~\ref{tab:simulation_results}, recovery depends on noise and not on sample size. The median shape RMSE is about $0.055$ with low noise, $0.09$ with medium noise and $0.13$ with high noise, with differences across sample sizes within each noise level smaller than the interquartile ranges, since each trajectory is reconstructed from the respondent's own thirteen judgments. The error of the first derivative follows the same pattern. Recovery differs across families (Table~\ref{tab:sim_family} in the Appendix): with low noise and $N=140$, the median shape RMSE is $0.029$ for linear, $0.049$ for concave, $0.067$ for sigmoid and $0.113$ for convex trajectories.

The centered generating trajectories have a low-dimensional structure, with $99.9\%$ of the variability explained by two components. The reconstructed trajectories remain low-dimensional, with $90$--$94\%$ explained by two components in the ordinary scenarios; the additional dimensions are due to reconstruction error. For group membership, the median ARI is between $0.46$ and $0.49$ with low noise, between $0.25$ and $0.28$ with medium noise and between $0.12$ and $0.13$ with high noise, and it does not increase with $N$. Continuous shape and low-dimensional structure are recovered more reliably than discrete group labels, and the clusters of Section~\ref{sec:results} summarize the dominant shape variation in the sample.

\section{Results}
\label{sec:results}

\subsection{Reconstructed trajectories}

\begin{figure}[htbp]
\centering
\includegraphics[width=\textwidth]{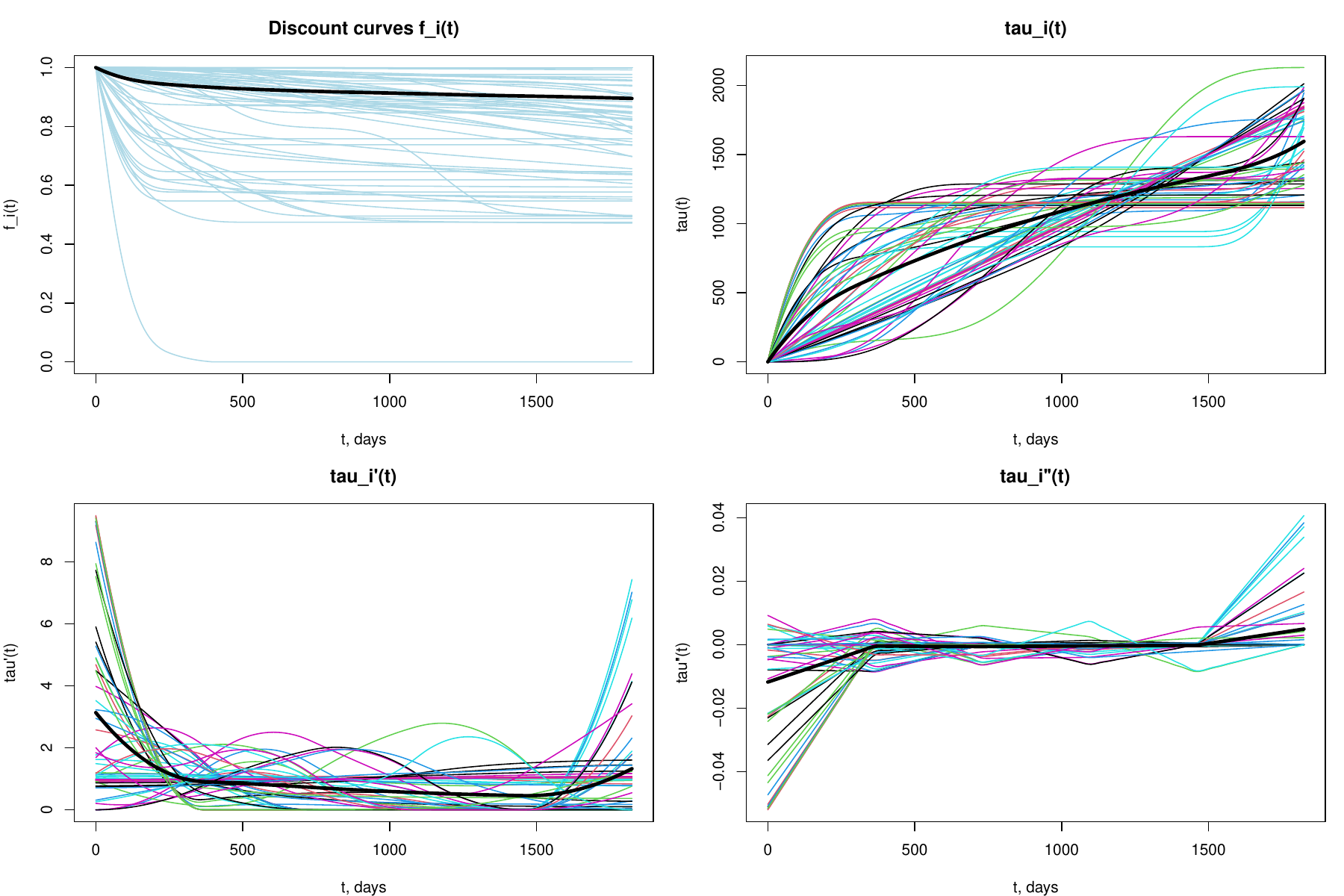}
\caption{Reconstruction for the $N=60$ identified respondents. Upper left: smoothed discount curves $\widehat f_i(t)$ (light lines) and their mean (thick line). Upper right: subjective-time trajectories $\tau_i(t)$, with the mean as a thick black line; a trajectory above the diagonal at horizon $t$ has accumulated more subjective time than calendar time up to $t$. Lower panels: first derivative $\tau_i'(t)$, the local speed of subjective time (values above one indicate faster-than-calendar accumulation), and second derivative $\tau_i''(t)$. Time in days.}
\label{fig:functional_reconstruction}
\end{figure}

Figure~\ref{fig:functional_reconstruction} shows the smoothed discount curves, the trajectories $\tau_i(t)$ and their derivatives for the $60$ respondents. The mean discount curve declines slowly, with a steeper decline in the first months. Individual curves range from almost flat profiles to curves that fall below $0.5$; one respondent has an equivalent that approaches zero within one year. Several curves become constant over the long horizons, reflecting identical equivalence judgments at two, three and five years. The individual scale $k_i$ has median $1.33\times10^{-4}$ per day (interquartile range $7.4\times10^{-5}$--$3.5\times10^{-4}$); the implied five-year discount factor $\exp(-1825k_i)$ ranges from $0.31$ to $0.999$, with one outlying value close to zero (Figure~\ref{fig:ki_dist} in the Appendix).

The trajectories differ in shape. Some rise steeply in the first months and then flatten, so that most subjective time is accumulated at short horizons; others stay close to or below the diagonal for one or two years and rise at long horizons. The first derivatives are highest at short horizons for the first group and at long horizons for the second, and the second derivatives show where the speed of subjective time changes. Trajectories of respondents with identical judgments over several long horizons are nearly constant over that range and increase between the three-year and the five-year horizon, where no intermediate horizon is observed.

\begin{figure}[htbp]
\centering
\includegraphics[width=0.85\textwidth]{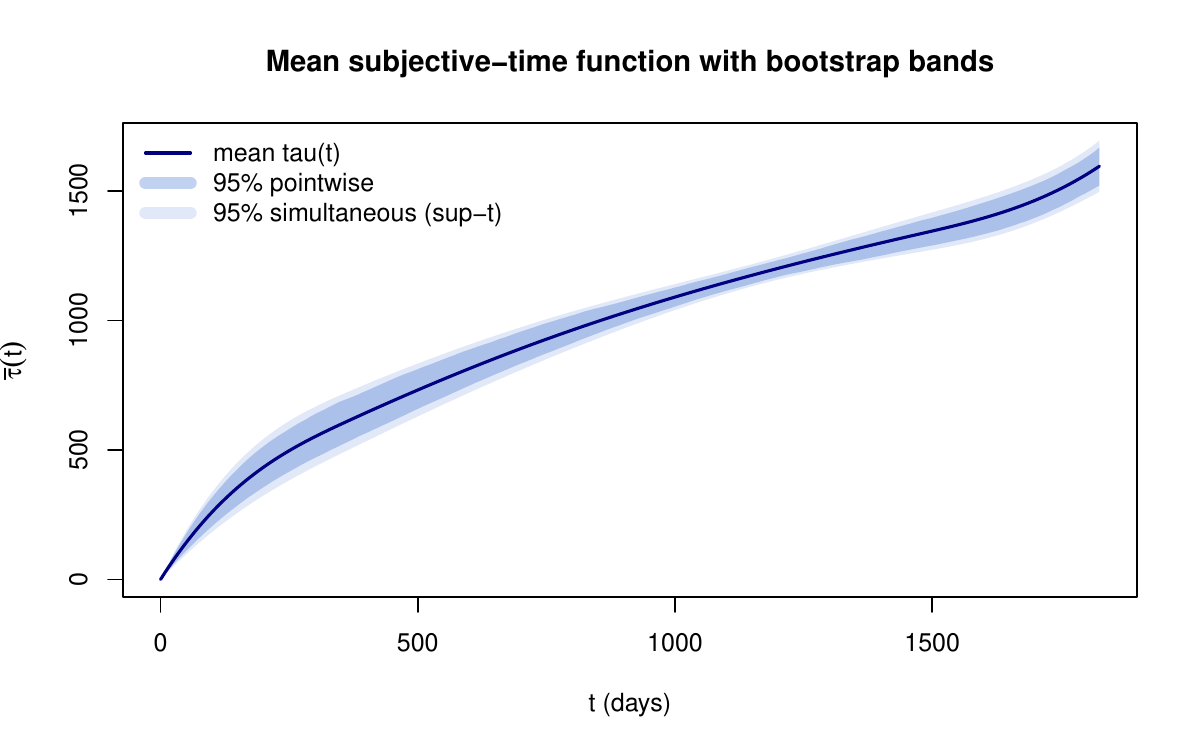}
\caption{Mean subjective-time trajectory $\bar\tau(t)$ for $N=60$ respondents with $95\%$ pointwise (inner) and simultaneous sup-$t$ (outer) bands from $B=1000$ subject-level bootstrap resamples. The bands describe sampling uncertainty of the mean, not the dispersion of individual trajectories. A concave mean indicates that, on average, subjective time accumulates faster than calendar time at short horizons and more slowly at long horizons.}
\label{fig:mean_tau_bootstrap}
\end{figure}

The mean trajectory (Figure~\ref{fig:mean_tau_bootstrap}) is predominantly concave. It equals $82$ subjective days at one month, $242$ at three months, $614$ at one year, $916$ at two years and $1143$ at three years, and reaches $1596$ at five years. Its average slope decreases from about $3$ subjective days per day over the first month to about $0.5$ between three and four years, and increases to about $0.7$ over the last year. The mean trajectory thus lies above the identity up to slightly beyond three years and below it afterwards, and its declining slope over the first four years is consistent with decreasing impatience for the average normalized trajectory. The simultaneous $95\%$ band (sup-$t$ critical value $2.658$) excludes the identity over most of the horizon: at one year it ranges from $502$ to $726$ days and at five years from $1497$ to $1695$ days.

\subsection{Parametric benchmarks}
\label{sec:benchmark_results}

All three parametric models converged for the $60$ respondents. Table~\ref{tab:parametric_benchmark} reports in-sample and leave-one-horizon-out errors. In sample, the exponential model has the largest errors (median RMSE $0.0253$) and is never the best-fitting model. The Weber--Fechner model has the lowest median RMSE ($0.0121$), followed by the FDA reconstruction ($0.0134$) and the Stevens model ($0.0144$); the FDA reconstruction has the lowest RMSE for $26$ respondents ($43.3\%$), the Stevens model for $18$ ($30.0\%$) and the Weber--Fechner model for $16$ ($26.7\%$). The mean RMSE of the FDA reconstruction ($0.0458$) is higher than those of the two subjective-time models ($0.0260$ and $0.0286$), because the distribution of its individual errors has a heavier right tail.

\begin{table}[htbp]
\centering
\caption{Parametric benchmarks ($N=60$). In-sample: errors at the thirteen horizons used for estimation. LOHO: each horizon is held out in turn, all models are re-estimated on the remaining twelve horizons and the held-out discount factor is predicted; RMSE is computed per respondent over the thirteen held-out predictions and summarized across respondents. Interior: held-out horizons from $14$ days to $3$ years. Best: share of respondents for whom the model has the lowest RMSE.}
\label{tab:parametric_benchmark}
\footnotesize
\setlength{\tabcolsep}{3pt}
\begin{tabular}{@{}lcccc@{}}
\toprule
\multicolumn{5}{@{}l}{\textit{In sample}}\\
Model & Median RMSE & Mean RMSE & Median MAE & Best (\%) \\
\midrule
Exponential & 0.0253 & 0.0745 & 0.0178 & 0.0 \\
Weber--Fechner & 0.0121 & 0.0260 & 0.0095 & 26.7 \\
Stevens & 0.0144 & 0.0286 & 0.0114 & 30.0 \\
FDA reconstruction & 0.0134 & 0.0458 & 0.0107 & 43.3 \\
\midrule
\multicolumn{5}{@{}l}{\textit{Leave-one-horizon-out}}\\
Model & Median RMSE [IQR] & Interior median RMSE & Median MAE & Best (\%) \\
\midrule
Exponential & 0.0312 [0.0094, 0.1185] & 0.0265 & 0.0216 & 20.0 \\
Weber--Fechner & 0.0225 [0.0087, 0.0426] & 0.0141 & 0.0156 & 30.0 \\
Stevens & 0.0264 [0.0079, 0.0559] & 0.0151 & 0.0162 & 30.0 \\
FDA reconstruction & 0.0261 [0.0096, 0.0628] & 0.0179 & 0.0198 & 20.0 \\
\bottomrule
\end{tabular}
\end{table}

When horizons are held out, the errors of all models increase and the differences between the two subjective-time models and the FDA reconstruction become small at the median: $0.0225$ for Weber--Fechner, $0.0261$ for FDA and $0.0264$ for Stevens, against $0.0312$ for the exponential model. The FDA reconstruction has a lower LOHO error than the exponential model for $56.7\%$ of respondents, and a lower error than the Weber--Fechner and Stevens models for $38.3\%$ and $36.7\%$, respectively. On interior horizons, the medians are $0.0141$ (Weber--Fechner), $0.0151$ (Stevens), $0.0179$ (FDA) and $0.0265$ (exponential). For $7$ of the $780$ held-out predictions, the FDA reconstruction without the held-out horizon did not determine $k_i$ and the smoothed discount curve was used. The errors by horizon are shown in Figure~\ref{fig:loho_horizon} in the Appendix. The two-parameter subjective-time models predict individual judgments accurately and parsimoniously, and the FDA reconstruction attains comparable median accuracy without assuming a parametric family. Figure~\ref{fig:parametric_benchmark_six_subjects} in the Appendix shows the four fits for six respondents.

\subsection{Functional principal components}

\begin{figure}[htbp]
\centering
\includegraphics[width=0.82\textwidth]{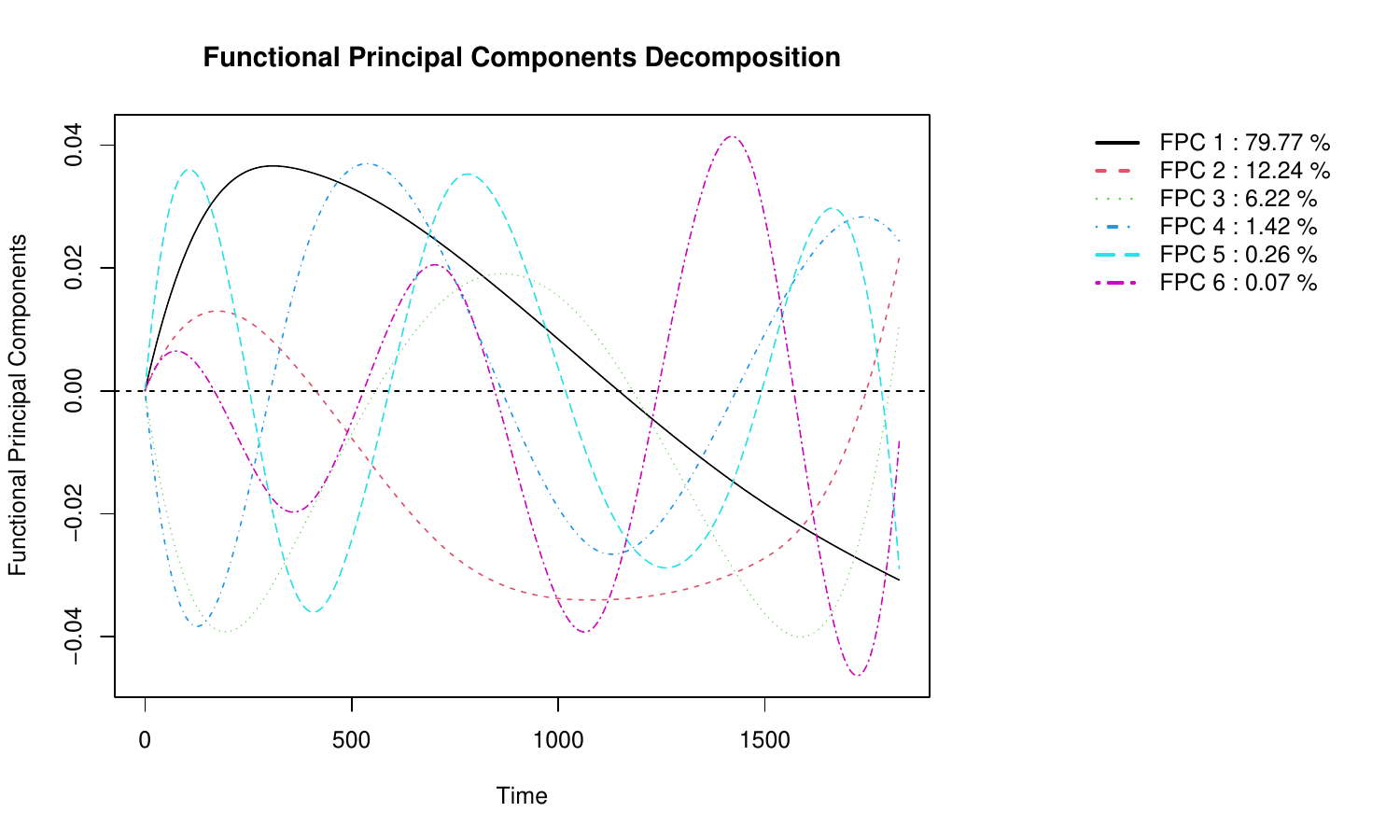}
\caption{First six eigenfunctions of the centered subjective-time trajectories $\tau_i^c(t)$, $N=60$, with the percentage of variability explained by each. The first component (black) contrasts accumulation of subjective time at short and medium horizons with accumulation at long horizons; the second (red, dashed) contrasts the central part of the horizon with its two ends. The sign of each eigenfunction is arbitrary.}
\label{fig:fpca_tau_centered}
\end{figure}

FPCA of the centered trajectories (Figure~\ref{fig:fpca_tau_centered}) shows a low-dimensional structure. The first component explains $79.77\%$ of the variability, the second $12.24\%$ and the third $6.22\%$; two components explain $92.00\%$, three $98.22\%$ and four $99.64\%$. The first eigenfunction is positive at short and medium horizons and negative beyond about three years: respondents with positive scores accumulate subjective time early and comparatively little later, and respondents with negative scores accumulate it late. The change of sign is consistent with the constraint \eqref{eq:orthogonality}; the component locates the change at about three years and measures its magnitude. The second eigenfunction contrasts the first year and the final months of the horizon with its central part. Perturbations of the mean along the two components are shown in Figure~\ref{fig:fpca_perturbation} in the Appendix. Following Section~\ref{sec:clustering}, the first two components are used for clustering.

The first two scores are weakly correlated with the scale $k_i$ ($r=0.127$ for FPC1 and $r=-0.300$ for FPC2), so that the dominant direction of shape variation is not a monotone function of discounting intensity.

\subsection{Profiles of subjective time}
\label{sec:cluster_results}

\begin{figure}[htbp]
\centering
\includegraphics[width=0.90\textwidth]{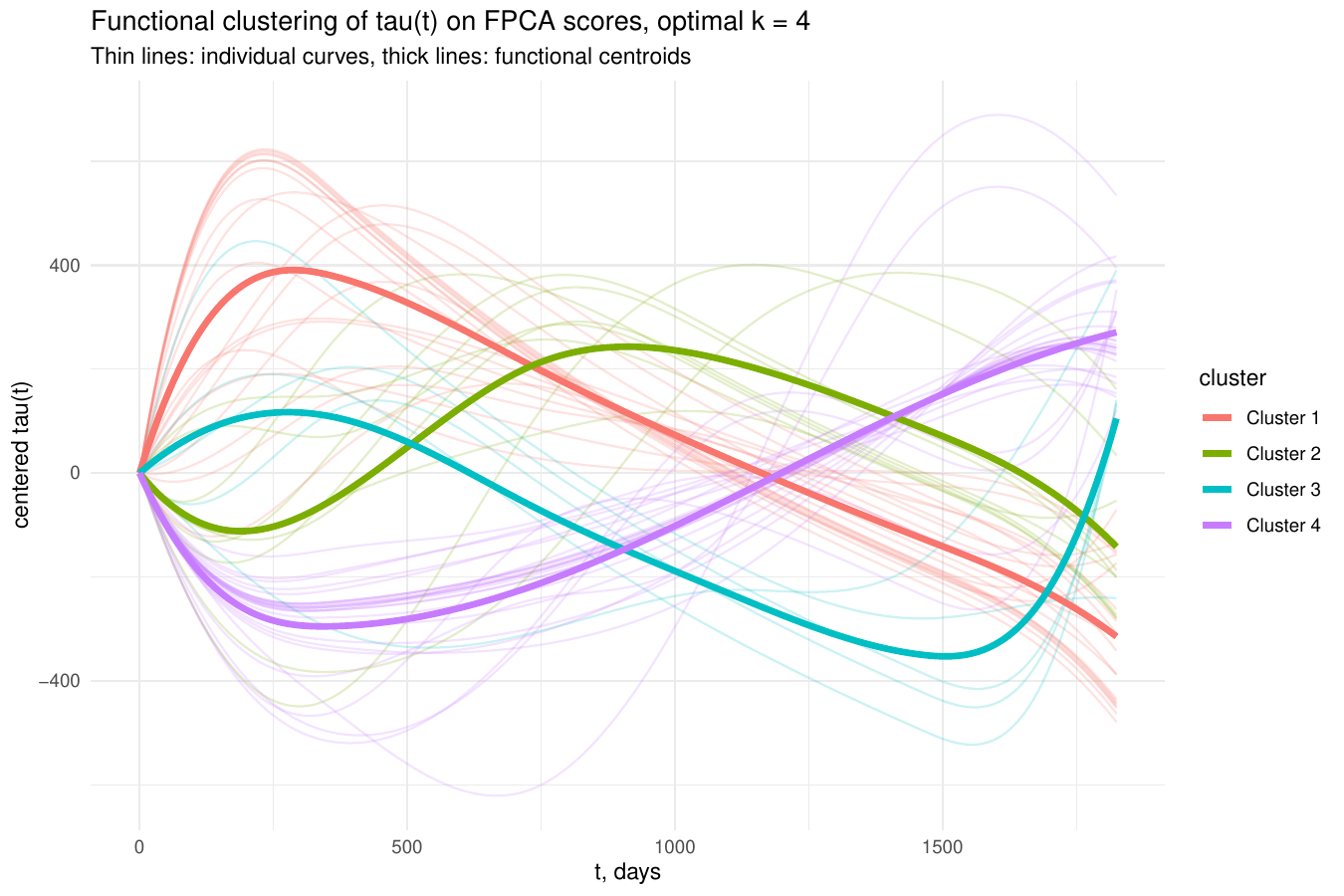}
\caption{Profiles of subjective time. Thin lines: centered trajectories $\tau_i^c(t)=\tau_i(t)-\bar\tau(t)$ of the $60$ respondents; thick lines: cluster centroids from $k$-means on the standardized scores of the first two FPCs ($K=4$, sizes $20/9/5/26$). Positive values indicate that, up to horizon $t$, members of the cluster have accumulated more subjective time than the sample mean; negative values indicate less. Time in days.}
\label{fig:functional_centroids_tau}
\end{figure}

The average silhouette is $0.468$ for $K=2$, $0.527$ for $K=3$, $0.569$ for $K=4$, $0.541$ for $K=5$ and $0.544$ for $K=6$; the four-cluster solution is adopted (Figure~\ref{fig:silhouette_selection} in the Appendix). The clusters contain $20$, $9$, $5$ and $26$ respondents. Their mean bootstrap Jaccard similarities are $0.891$, $0.818$, $0.910$ and $0.992$, and the clusters dissolve (Jaccard below $0.5$) in $27$, $45$, $56$ and $0$ of the $500$ resamples, respectively. Figure~\ref{fig:functional_centroids_tau} shows individual centered trajectories and cluster centroids.

The profiles differ in when subjective time accumulates relative to the mean. Cluster~1 ($n=20$) accumulates subjective time early: its centroid rises to $+391$ subjective days at about $280$ days, crosses zero at about $1160$ days and ends at $-315$ at five years; all individual trajectories in this cluster end below the mean. Cluster~4 ($n=26$) shows the opposite pattern: its centroid falls to $-295$ at about $340$ days, crosses zero at about $1200$ days and ends at $+271$, and all its members end above the mean. Clusters~1 and~4, which together include $46$ of the $60$ respondents, correspond to the two ends of the first component. Cluster~2 ($n=9$) is close to the mean at short horizons (minimum $-112$ at about $190$ days), accumulates more subjective time than the mean in the central part of the horizon (maximum $+243$ at about $910$ days) and ends at $-142$. Cluster~3 ($n=5$) has a moderate early excess (maximum $+117$ at about $275$ days), a deficit at long horizons (minimum $-353$ at about $1500$ days) and a final value of $+105$; the increase after the minimum takes place between the three-year and the five-year horizons.

The profiles differ in discounting intensity. The median $k_i$ is $3.7\times10^{-4}$ in Cluster~1, $1.0\times10^{-4}$ in Cluster~2, $1.4\times10^{-4}$ in Cluster~3 and $8.0\times10^{-5}$ in Cluster~4; the corresponding implied five-year discount factors are $0.51$, $0.83$, $0.77$ and $0.86$. The difference is detected by an ANOVA on $\log k_i$ ($F(3,56)=7.86$, $p<0.001$) and by a Kruskal--Wallis test ($\chi^2(3)=22.50$, $p<0.001$), while the ANOVA on the raw scale, which is dominated by the right tail of $k_i$, gives $p=0.084$. Early accumulation of subjective time is thus more frequent among stronger discounters and late accumulation among weaker discounters, although the correlations with the FPC scores show that shape and intensity are not equivalent.

The covariates not used in clustering are described in Table~\ref{tab:cluster_profile}. Mean age ranges from $38.1$ to $46.2$ years and mean TSS$_{\text{core}}$ from $4.78$ to $5.40$ across clusters. Mean financial literacy is higher in Clusters~2 ($2.67$) and~3 ($2.40$) than in Clusters~1 ($1.70$) and~4 ($1.85$), and the share of men is $88.9\%$ in Cluster~2 and between $53.8\%$ and $65.0\%$ in the other clusters. These comparisons are descriptive.

\begin{table}[htbp]
\centering
\caption{Profiling of the four clusters. $k_i$ is estimated from the same judgments as the trajectories; the other variables are not used in the construction of the trajectories. Mean (SD) unless otherwise stated. $k_i$: median per day.}
\label{tab:cluster_profile}
\footnotesize
\begin{tabular}{lcccc}
\toprule
 & Cluster 1 & Cluster 2 & Cluster 3 & Cluster 4 \\
 & ($n=20$) & ($n=9$) & ($n=5$) & ($n=26$) \\
\midrule
$k_i$ (median) & $3.7\times10^{-4}$ & $1.0\times10^{-4}$ & $1.4\times10^{-4}$ & $8.0\times10^{-5}$ \\
Age (years) & 41.5 (11.6) & 38.1 (13.0) & 46.2 (19.2) & 40.7 (13.7) \\
Men, $n$ (\%) & 13 (65.0) & 8 (88.9) & 3 (60.0) & 14 (53.8) \\
TSS$_{\text{core}}$ & 4.78 (1.24) & 4.91 (1.15) & 5.40 (0.66) & 4.86 (1.13) \\
Financial literacy & 1.70 (1.03) & 2.67 (0.71) & 2.40 (0.55) & 1.85 (0.88) \\
Mean STP & 3.12 (3.34) & 4.53 (3.38) & 1.62 (1.60) & 2.68 (2.98) \\
\bottomrule
\end{tabular}
\end{table}

Clustering of the parametric parameters does not reproduce the functional partition (Table~\ref{tab:parametric_clustering_ari}). The ARI is $0.286$ for the Weber--Fechner parameters including $k_i$ and $0.004$ without it, $-0.003$ for the Stevens parameters including $k_i$ and $0.260$ without it, and $0.258$ with PAM on the full Stevens space.

\begin{table}[htbp]
\centering
\caption{Agreement (ARI) between the functional partition ($K=4$, $N=60$) and partitions obtained by clustering standardized parametric estimates into four groups.}
\label{tab:parametric_clustering_ari}
\footnotesize
\begin{tabular}{lcc}
\toprule
Parametric space & Algorithm & ARI \\
\midrule
Weber--Fechner $(k_i,\alpha_i,\beta_i)$ & $k$-means & 0.286 \\
Weber--Fechner $(\alpha_i,\beta_i)$ & $k$-means & 0.004 \\
Stevens $(k_i,\alpha_i,\beta_i)$ & $k$-means & $-0.003$ \\
Stevens $(\alpha_i,\beta_i)$ & $k$-means & 0.260 \\
Stevens $(k_i,\alpha_i,\beta_i)$ & PAM & 0.258 \\
\bottomrule
\end{tabular}
\end{table}

\subsection{Implicit and explicit subjective time}
\label{sec:stp_results}

\begin{figure}[htbp]
\centering
\includegraphics[width=\textwidth]{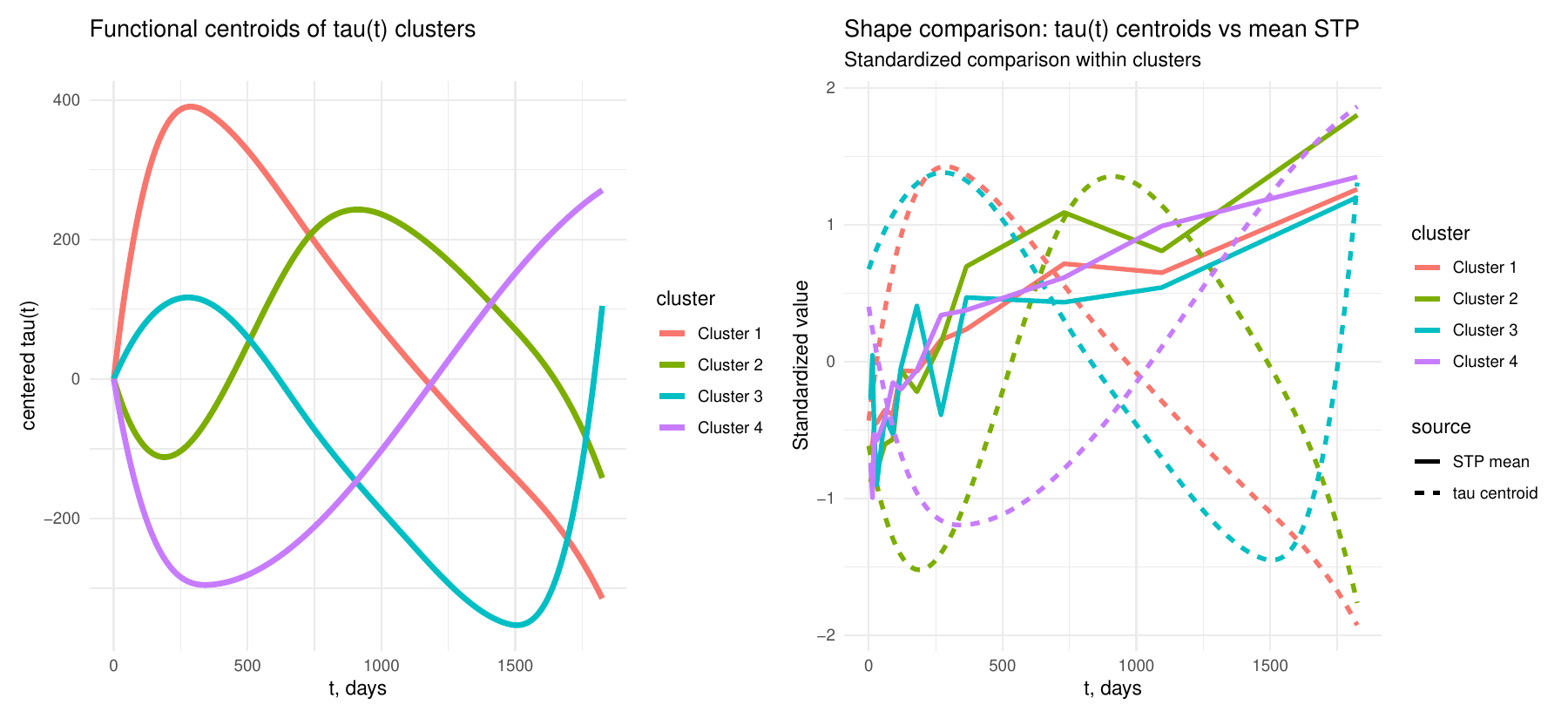}
\caption{Implicit and explicit subjective time by cluster. Left: centroids of the centered trajectories $\tau_i^c(t)$. Right: standardized centroid of $\tau$ (dashed) and standardized mean STP profile (solid) for each cluster at the thirteen horizons. Similar shapes of dashed and solid lines of the same color would indicate agreement between the two measures within a cluster.}
\label{fig:tau_stp_comparison}
\end{figure}

Figure~\ref{fig:tau_stp_comparison} compares the centroids of $\tau$ with the cluster means of the standardized STP profiles. The STP profiles increase with the horizon in all four clusters and are similar across clusters, whereas the $\tau$ centroids differ in the timing of their deviations from the mean.

The permutation tests do not detect a relation between the two measures (Table~\ref{tab:stp_tests}). The STP profiles do not differ across clusters, either in shape (T1, interaction $F=1.09$, $p=0.417$) or in level ($F=1.17$, $p=0.337$). The correlations between centered $\tau$ centroids and centered STP means are $-0.33$, $0.32$, $0.32$ and $-0.20$ in the four clusters, all within the central $95\%$ of their permutation distributions, and the global statistic $G=0.076$ has $p=0.604$. At the scalar level, mean STP is weakly correlated with $k_i$ ($r=0.258$), FPC1 ($r=-0.170$), FPC2 ($r=-0.258$) and TSS$_{\text{core}}$ ($r=0.109$), and $k_i$ and TSS$_{\text{core}}$ are nearly uncorrelated ($r=-0.087$). No statistically detectable correspondence between choice-implied trajectories and explicit positioning profiles is observed in the analytical sample.

\begin{table}[htbp]
\centering
\caption{Permutation tests for the relation between implicit trajectories and explicit STP profiles ($N=60$, $B=5000$). T1: cluster differences in STP profiles (labels permuted). T2: agreement between $\tau$ centroids and STP means (STP profiles permuted across respondents); $r_k$ computed over the twelve horizons from $14$ days to $5$ years.}
\label{tab:stp_tests}
\footnotesize
\begin{tabular}{lcccc}
\toprule
Statistic & $n$ & Observed & Null $95\%$ range & $p$ \\
\midrule
T1: cluster $\times$ horizon interaction ($F$) & 60 & 1.086 & -- & 0.417 \\
T1: respondent mean STP ($F$) & 60 & 1.169 & -- & 0.337 \\
T2: $r_1$ & 20 & $-0.326$ & $[-0.584, 0.573]$ & 0.438 \\
T2: $r_2$ & 9 & $0.315$ & $[-0.558, 0.592]$ & 0.441 \\
T2: $r_3$ & 5 & $0.315$ & $[-0.568, 0.554]$ & 0.302 \\
T2: $r_4$ & 26 & $-0.200$ & $[-0.536, 0.540]$ & 0.618 \\
T2: $G$ & 60 & 0.076 & $[0.017, 0.263]$ & 0.604 \\
\bottomrule
\end{tabular}
\end{table}

\subsection{Sensitivity of the profiles}
\label{sec:sensitivity_results}

Table~\ref{tab:robustness} summarizes the agreement between the reference partition and the alternatives of Section~\ref{sec:sensitivity}. The partition is identical across $k$-means variants and is close to the PAM ($0.929$) and Ward.D2 ($0.860$) solutions on the same scores. It is also unchanged when the three trajectories with the largest curvature are removed: the reduced sample of $57$ respondents gives clusters of sizes $17/9/5/26$, ARI $1.000$ on the common respondents, two components explaining $91.6\%$ of the variability and a relative $L^2$ shift of the mean of $0.017$. Adding components lowers the silhouette and the agreement (ARI $0.755$ with three and $0.614$ with four components). Agreement is $0.656$ and $0.664$ with the $L^2$ distance between centered trajectories, $0.51$--$0.55$ with the first-derivative semimetric and $0.24$--$0.30$ with the second-derivative semimetric, which groups respondents by local curvature and not by the timing of deviations. Across the nine basis combinations, the variance explained by two components ranges from $89.2\%$ to $97.9\%$, and the ARI of the eight alternative combinations ranges from $0.605$ to $1.000$. With the ratio-of-sums rule for $k_i$, the two scales correlate at $0.996$, two components explain $97.4\%$ of the variability, the cluster sizes are $18/12/6/24$ and the ARI is $0.589$. Over the $26$ alternative specifications of Table~\ref{tab:robustness}, the median ARI is $0.710$ and $10$ have ARI of at least $0.80$. The contrast between early and late accumulation, which defines the first component and the two largest clusters, is the most stable feature of the partition; the boundaries of the two smaller clusters depend more on the specification.

\begin{table}[htbp]
\centering
\caption{Agreement (ARI) between the reference partition ($K=4$, standardized scores of two FPCs, $k$-means) and alternative specifications, $N=60$.}
\label{tab:robustness}
\footnotesize
\begin{tabular}{@{}p{0.68\textwidth}c@{}}
\toprule
Specification & ARI \\
\midrule
$k$-means, $25$/$50$/$100$ starts; Hartigan--Wong, Lloyd, MacQueen & 1.000 \\
PAM / Ward.D2 on standardized scores & 0.929 / 0.860 \\
Three / four / five / six FPCs & 0.755 / 0.614 / 0.465 / 0.608 \\
$L^2$ distance on $\tau_i^c$, PAM / Ward.D2 & 0.656 / 0.664 \\
First-derivative semimetric, PAM / Ward.D2 & 0.552 / 0.509 \\
Second-derivative semimetric, PAM / Ward.D2 & 0.297 / 0.241 \\
Basis dimensions ($f$: 8--12; $\tau$: 6--10), range over eight alternative combinations & 0.605--1.000 \\
Ratio-of-sums rule for $k_i$ & 0.589 \\
Excluding the three highest-curvature trajectories ($N=57$) & 1.000 \\
\bottomrule
\end{tabular}
\end{table}

\section{Discussion and conclusions}
\label{sec:discussion}

This paper represented intertemporal preferences through individual subjective-time trajectories reconstructed from equivalence judgments, and analyzed them as functional data. The representation $f_i(t)=\exp\{-k_i\tau_i(t)\}$ determines only the product of scale and trajectory, and the operational normalization adopted here makes $\tau_i(t)$ coincide with chronological time for an exponential discounter. Within this normalization, the shape of $\tau_i(t)$ describes where along the horizon each respondent's impatience is concentrated.

On average, the trajectories are predominantly concave: subjective time accumulates faster than calendar time up to about one year and more slowly afterwards, and the simultaneous bootstrap band excludes the identity over most of the horizon. For the average normalized trajectory, this pattern is consistent with decreasing impatience, with the concave transformations reported by direct measurements of perceived duration \citep{Zauberman_2009} and with the link between concave time distortions and log-convex discount functions \citep{Prelec_2004,Cruz_Rambaud_2023}. Around this mean, individual variation is concentrated on very few directions: one functional component, contrasting early and late accumulation of subjective time, explains about four fifths of the variability, and two components explain $92\%$. The four profiles derived from these components characterize the analytical sample, in which most respondents either accumulate subjective time early (Cluster~1) or late (Cluster~4), with two smaller groups whose deviations are concentrated in the central part of the horizon or at its end. The profiles are associated with discounting intensity, early accumulation being more common among stronger discounters, but the weak correlations between $k_i$ and the component scores and the low agreement between the functional partition and the parametric partitions indicate that the timing of impatience is a dimension of heterogeneity that a scale parameter, or the two parameters of a Weber--Fechner or Stevens transformation, do not summarize.

The Weber--Fechner and Stevens models fit individual judgments well and predict held-out horizons with median errors comparable to, or lower than, those of the functional reconstruction, while the exponential model in chronological time is the least accurate. Parametric models summarize each respondent through interpretable parameters; the functional reconstruction provides a common representation, without a fixed parametric family, in which trajectories of different shapes are decomposed, compared and grouped across respondents.

The implicit trajectories and the explicit positioning task do not show detectable agreement. Explicit STP profiles increase smoothly with the horizon and are similar across clusters, and neither the cluster-level shapes nor their pairing with individual trajectories differ from what is expected under permutation. Together with the weak correlations among $k_i$, TSS$_{\text{core}}$ and mean STP, no statistically detectable correspondence between subjective time inferred from choices and subjective time stated in a perception task emerges in this sample, a result compatible with evidence that discounting and explicit time perception can respond differently to experimental manipulations \citep{Xu_2020}.

The simulation study indicates that the procedure recovers the shape of individual trajectories with an accuracy governed by response noise and that the reconstructed trajectories retain the low-dimensional structure of the generating process, while the recovery of group membership depends mainly on noise. In the empirical application, the contrast between early and late accumulation is the most stable feature across specifications, and the four-cluster solution describes trajectory heterogeneity under the reference specification.

The procedure applies to any design that elicits equivalence judgments over a common set of horizons. A new respondent's trajectory can be reconstructed with the same smoothing and normalization, centered on the reference mean, projected on the reference eigenfunctions, standardized with the reference score moments and assigned to the nearest reference centroid. Further developments include designs with more observations at long horizons and functional regression models relating $\tau_i(t)$ to individual characteristics.

\backmatter

\bmhead{Acknowledgements}
Not applicable.

\section*{Declarations}

\begin{itemize}
\item \textbf{Funding.} The authors declare that no funds, grants, or other support were received during the preparation of this manuscript.
\item \textbf{Competing interests.} The authors have no relevant financial or non-financial interests to disclose.
\item \textbf{Consent to participate.} Participation was voluntary and uncompensated. Before beginning the questionnaire, participants were informed of the scientific purpose of the study, the voluntary nature of participation, their right to discontinue at any time, and the anonymous and aggregate use of their responses for scientific research. No directly identifying information was collected. Participants provided informed consent by proceeding with the questionnaire after receiving this information. Data processing complied with Regulation (EU) 2016/679 (GDPR).
\item \textbf{Data availability.} The data were collected through an anonymous online questionnaire and are not publicly available for privacy reasons. An anonymized version of the dataset may be made available by the corresponding author upon reasonable request, subject to applicable data protection requirements.
\item \textbf{Code availability.} The R code used for the analyses is available from the corresponding author upon reasonable request.
\item \textbf{Use of generative AI.} The authors used generative AI tools to improve the English language, to assist in revising parts of the text, and to check bibliographic metadata. All content was critically reviewed, verified and edited by the authors, who take full responsibility for the publication.
\end{itemize}

\begin{appendices}

\section{Descriptive distributions}

\begin{figure}[htbp]
\centering
\includegraphics[width=\textwidth]{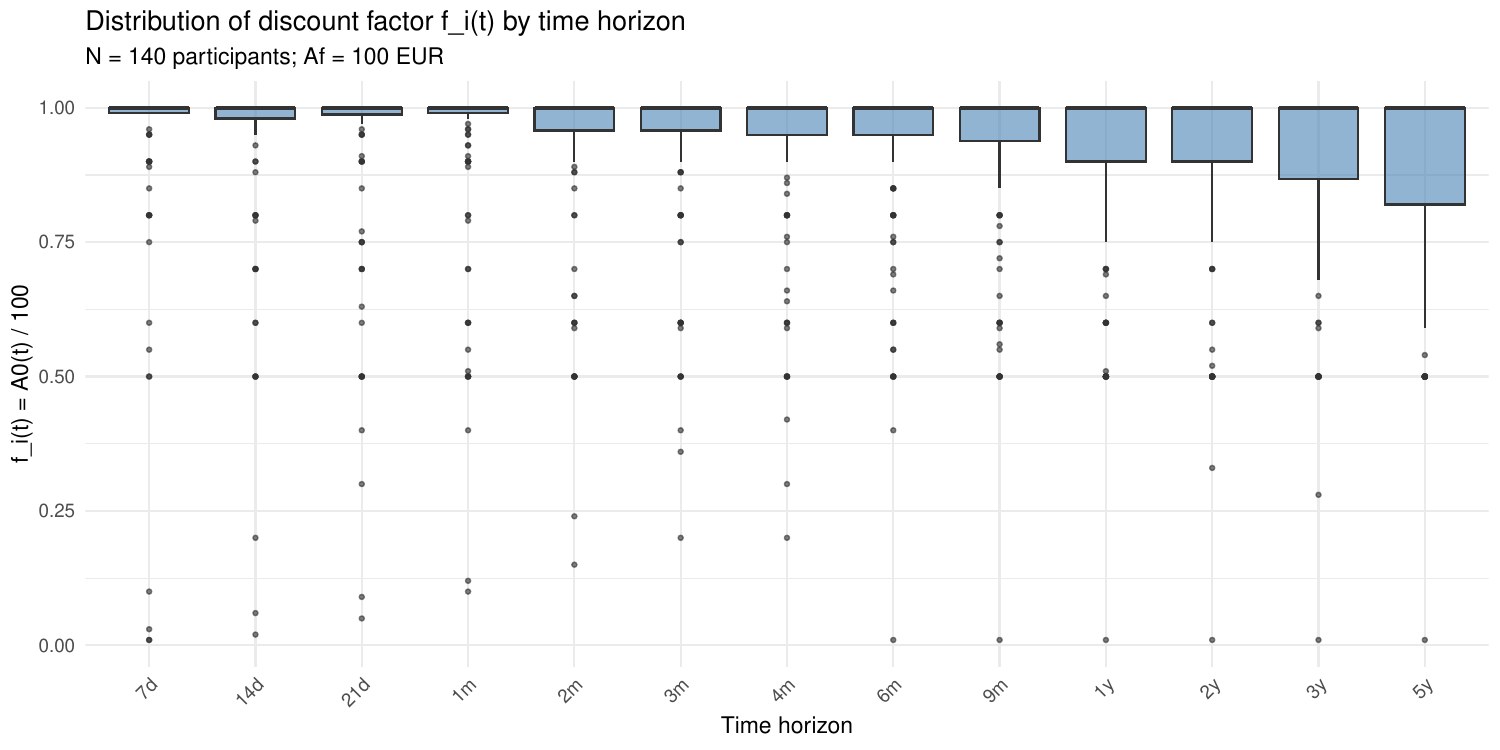}
\caption{Observed discount factors $f_i(t)=A_{0,i}(t)/100$ by horizon for the $N=140$ eligible respondents. Boxes show quartiles and medians. The median equals one at every horizon: between $76$ and $93$ respondents report the full future amount at each horizon, $61$ of them at all horizons.}
\label{fig:eq_dist}
\end{figure}

\begin{figure}[htbp]
\centering
\includegraphics[width=0.75\textwidth]{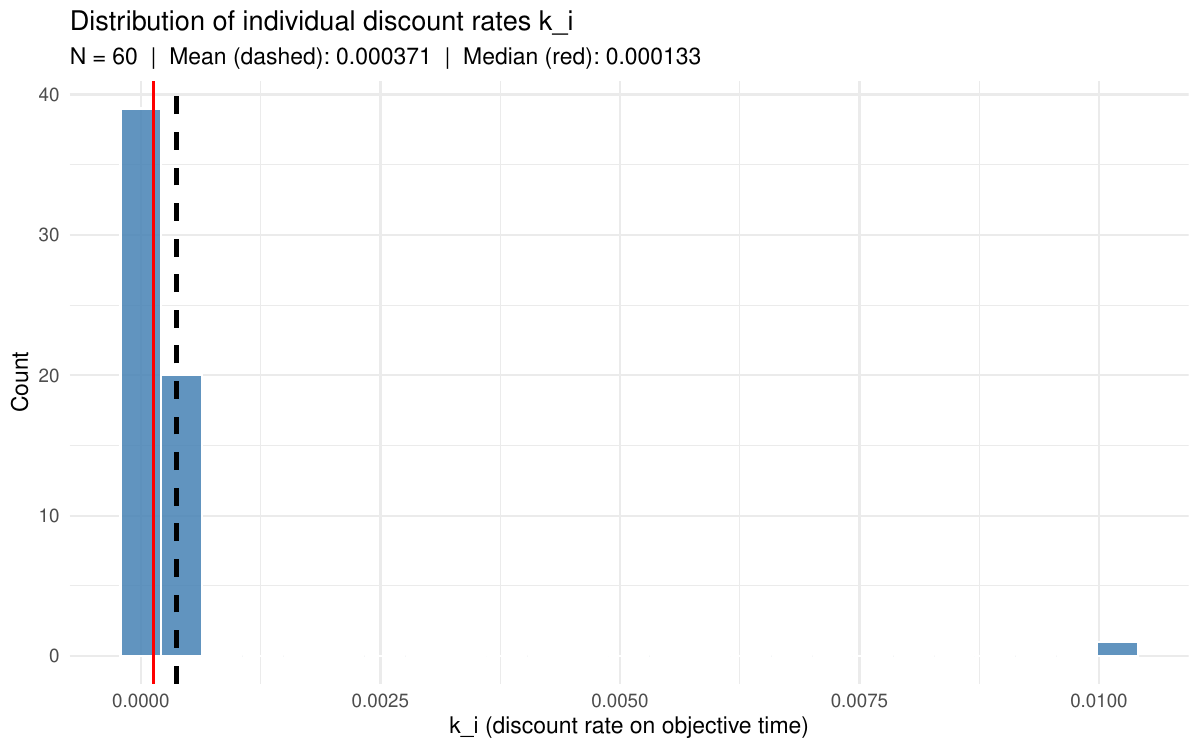}
\caption{Distribution of the operational discounting scale $k_i$ (per day) for the $N=60$ identified respondents. The distribution is right-skewed; one respondent has $k_i=0.0102$, corresponding to an implied five-year discount factor close to zero.}
\label{fig:ki_dist}
\end{figure}

\clearpage
\section{Simulation study by generating family}

\begin{table}[htbp]
\centering
\caption{Monte Carlo recovery by generating family for $N=140$ and for the stress scenario: medians over $R=100$ replications. $\bar n$: mean number of simulated respondents per family. Shape RMSE as in Table~\ref{tab:simulation_results}.}
\label{tab:sim_family}
\footnotesize
\begin{tabular}{llccc}
\toprule
Scenario & Family & $\bar n$ & Shape RMSE & RMSE $\tau'(t)$ \\
\midrule
\multirow{4}{*}{$N=140$, low} & Linear & 35.3 & 0.0286 & $5.9\times10^{-5}$ \\
 & Concave & 34.9 & 0.0486 & $2.2\times10^{-4}$ \\
 & Convex & 34.7 & 0.1132 & $4.0\times10^{-4}$ \\
 & Sigmoid & 35.0 & 0.0665 & $2.9\times10^{-4}$ \\
\midrule
\multirow{4}{*}{$N=140$, medium} & Linear & 34.8 & 0.0494 & $1.0\times10^{-4}$ \\
 & Concave & 35.2 & 0.0892 & $3.3\times10^{-4}$ \\
 & Convex & 35.8 & 0.1320 & $4.2\times10^{-4}$ \\
 & Sigmoid & 34.1 & 0.0987 & $3.6\times10^{-4}$ \\
\midrule
\multirow{4}{*}{$N=140$, high} & Linear & 35.5 & 0.0803 & $1.7\times10^{-4}$ \\
 & Concave & 35.0 & 0.1473 & $4.5\times10^{-4}$ \\
 & Convex & 34.7 & 0.1596 & $4.3\times10^{-4}$ \\
 & Sigmoid & 34.7 & 0.1307 & $4.6\times10^{-4}$ \\
\midrule
\multirow{4}{*}{Stress} & Linear & 24.5 & 0.0724 & $1.5\times10^{-4}$ \\
 & Concave & 24.7 & 0.1691 & $4.4\times10^{-4}$ \\
 & Convex & 25.4 & 0.1714 & $3.8\times10^{-4}$ \\
 & Sigmoid & 25.4 & 0.1351 & $4.6\times10^{-4}$ \\
\bottomrule
\end{tabular}
\end{table}

\begin{figure}[htbp]
\centering
\includegraphics[width=0.85\textwidth]{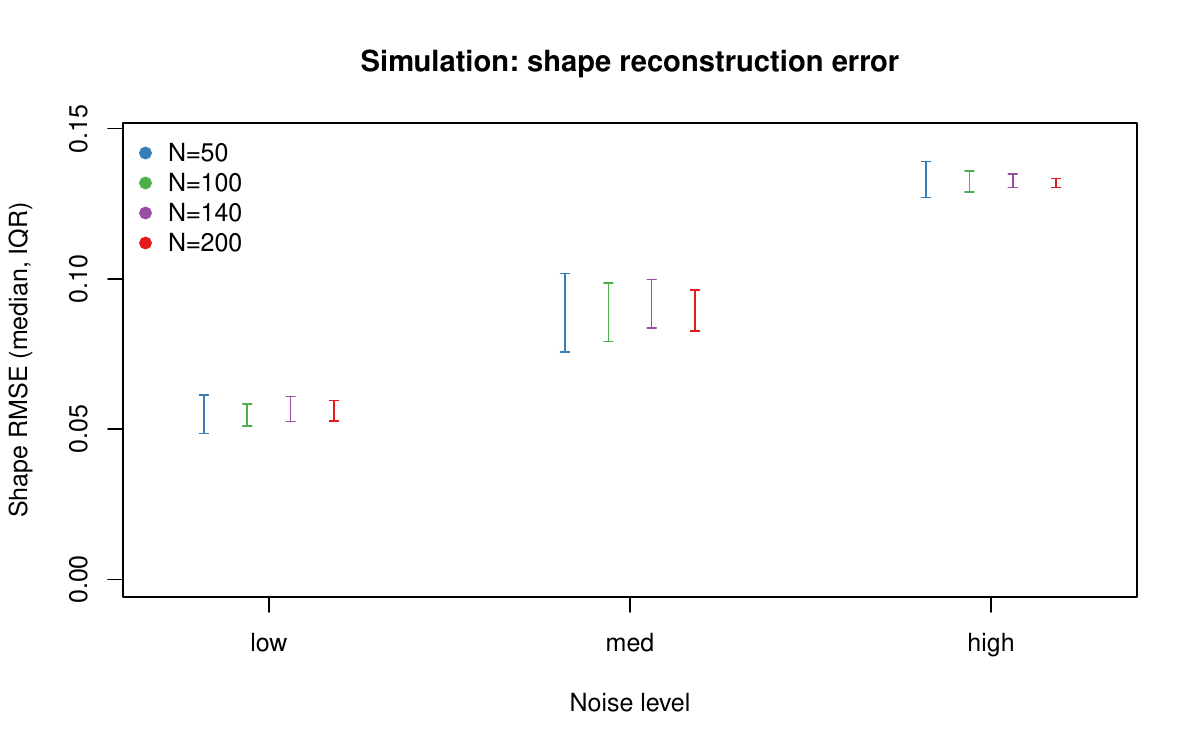}
\caption{Monte Carlo distribution of the shape RMSE between generating and reconstructed trajectories by scenario ($R=100$ replications per scenario). Recovery deteriorates with noise and changes little with sample size.}
\label{fig:sim_rmse}
\end{figure}

\begin{figure}[htbp]
\centering
\includegraphics[width=0.85\textwidth]{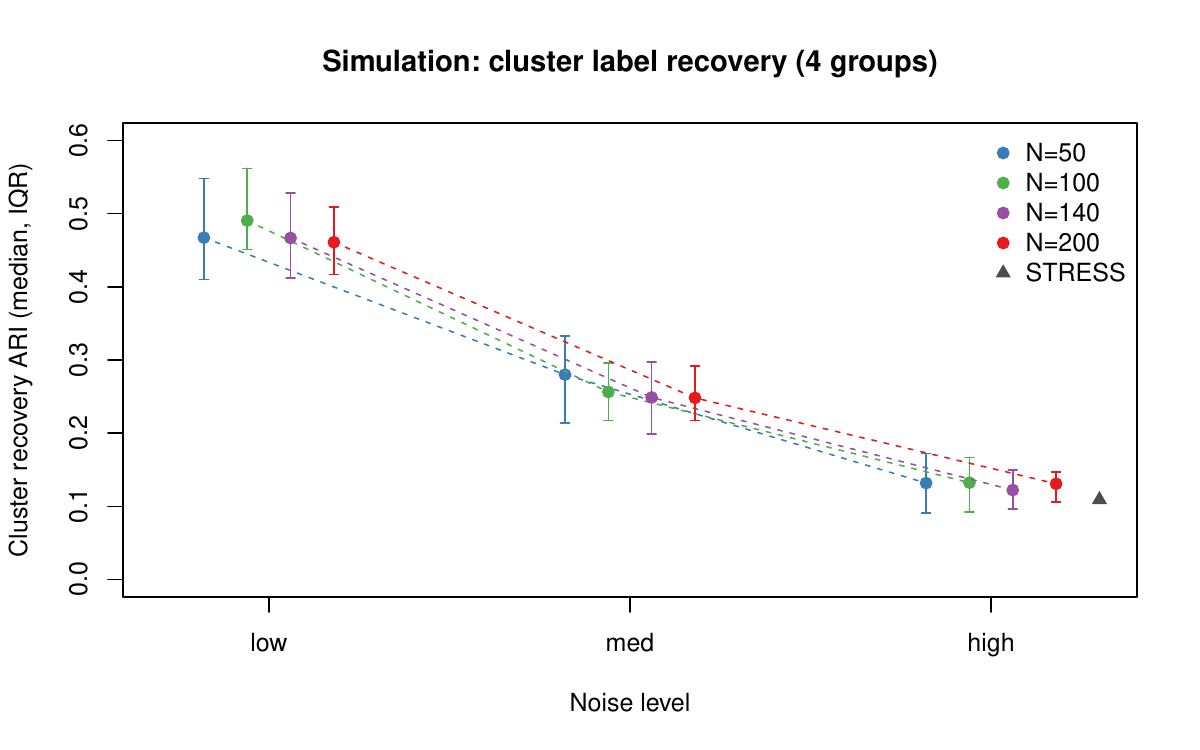}
\caption{Monte Carlo distribution of the Adjusted Rand Index between the four generating families and the four clusters estimated from the reconstructed trajectories, by scenario. Values close to one would indicate full recovery of group membership.}
\label{fig:sim_ari}
\end{figure}

\clearpage
\section{Parametric benchmarks}

\begin{figure}[htbp]
\centering
\includegraphics[width=\textwidth]{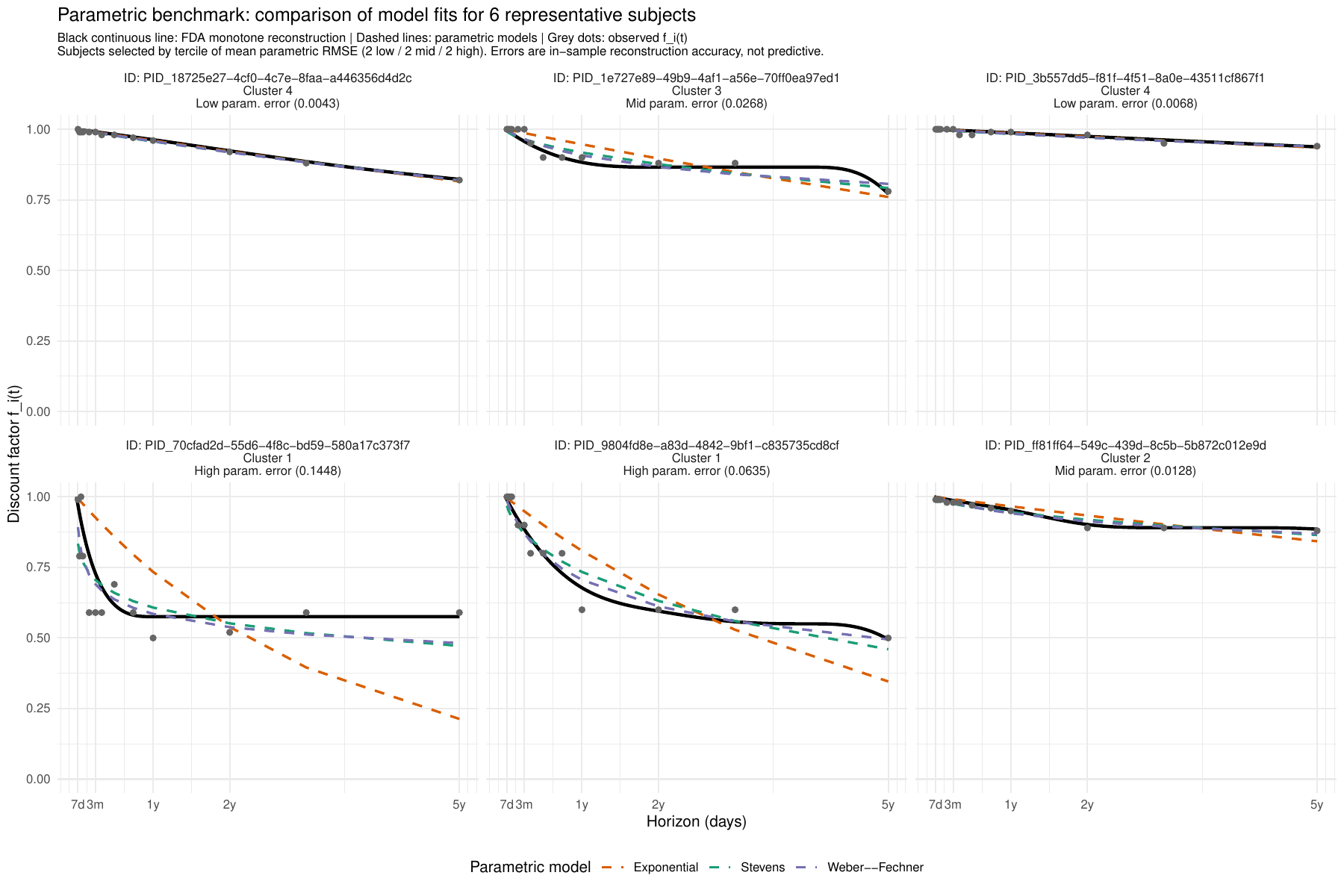}
\caption{Observed discount factors (points), exponential, Weber--Fechner and Stevens fits, and FDA reconstruction $\exp\{-k_i\tau_i(t)\}$ for six respondents selected by tertiles of the average parametric RMSE. The panels show cases in which the parametric forms and the functional reconstruction describe the data similarly and cases in which they differ.}
\label{fig:parametric_benchmark_six_subjects}
\end{figure}

\begin{figure}[htbp]
\centering
\includegraphics[width=0.85\textwidth]{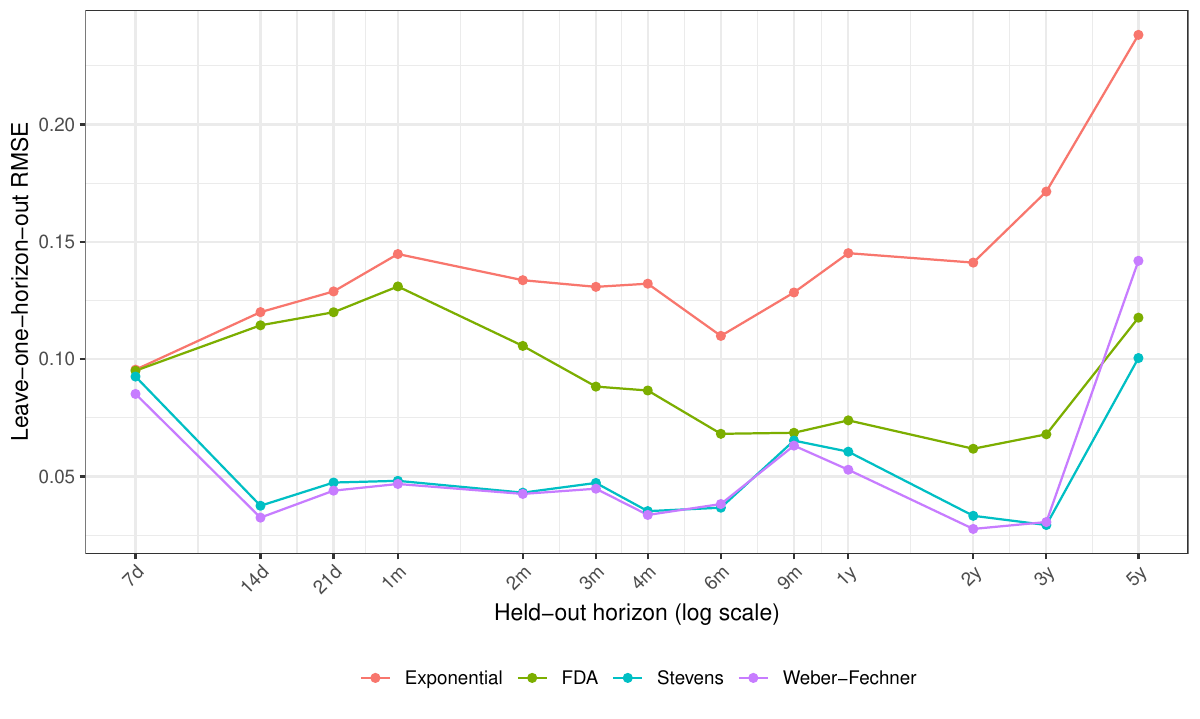}
\caption{Leave-one-horizon-out RMSE by held-out horizon, pooled over the $60$ respondents. Each point is the root mean squared prediction error for the discount factor at that horizon when it is excluded from estimation. Pooled errors are sensitive to a few respondents with large prediction errors; per-respondent medians are reported in Table~\ref{tab:parametric_benchmark}.}
\label{fig:loho_horizon}
\end{figure}

\clearpage
\section{Functional principal components and clustering}

\begin{figure}[htbp]
\centering
\includegraphics[width=0.48\textwidth]{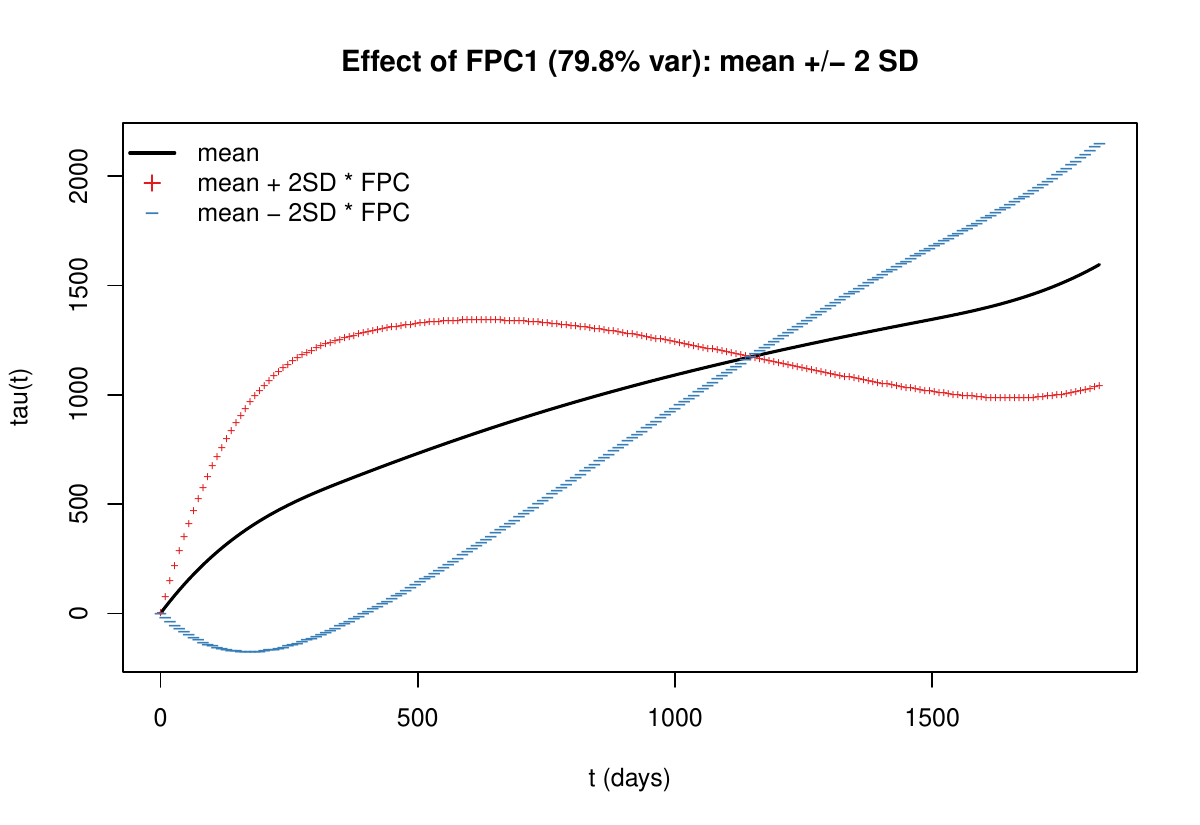}\hfill
\includegraphics[width=0.48\textwidth]{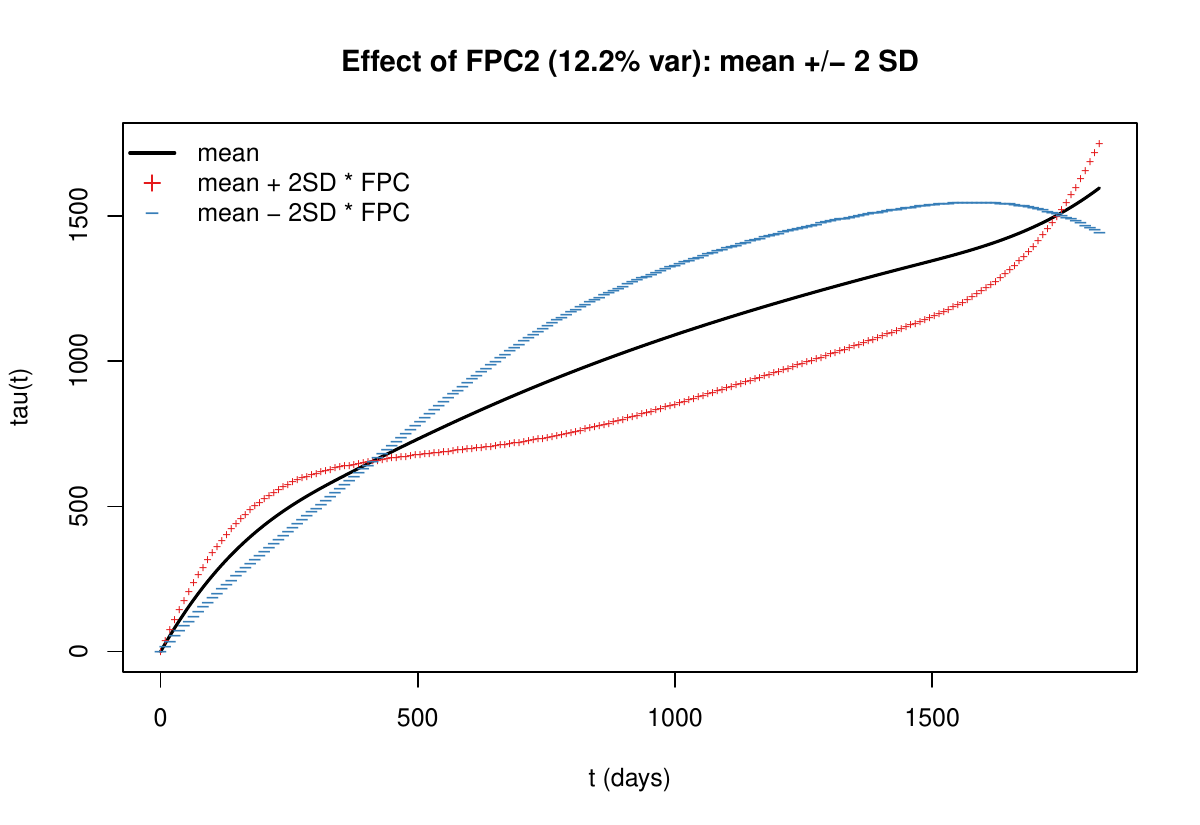}
\caption{Mean trajectory $\bar\tau(t)$ and perturbations along the first (left) and second (right) functional principal components. The perturbed curves show the modes of variation associated with high and low scores on each component; they are not necessarily admissible individual trajectories.}
\label{fig:fpca_perturbation}
\end{figure}

\begin{figure}[htbp]
\centering
\includegraphics[width=0.70\textwidth]{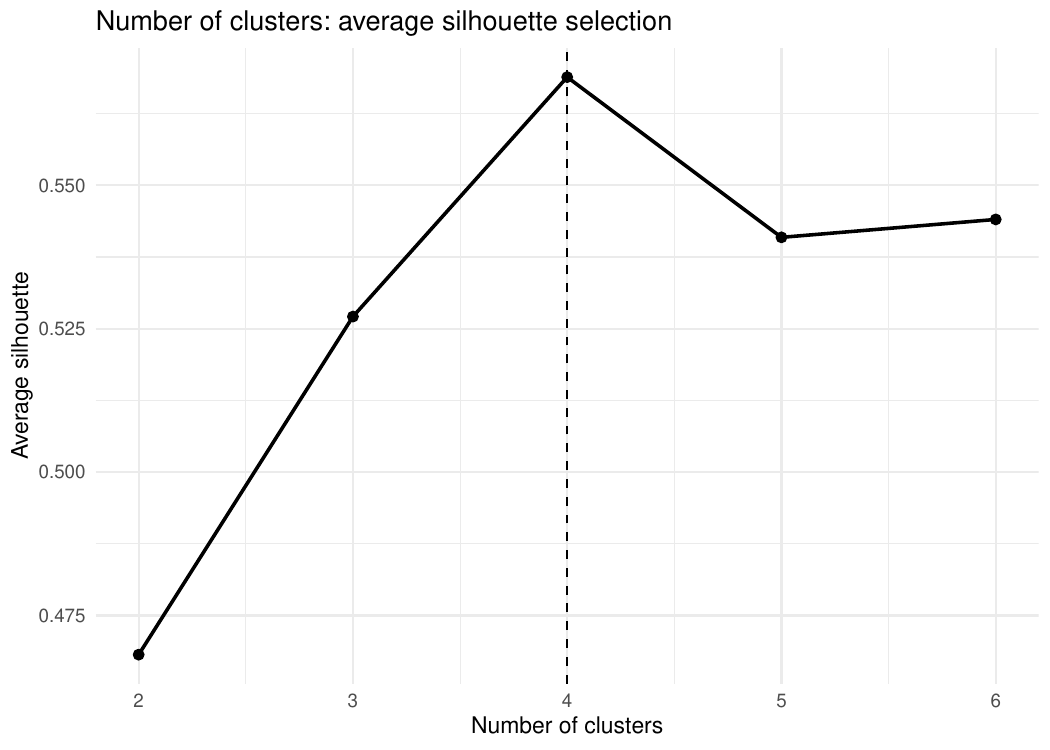}
\caption{Average silhouette of $k$-means partitions of the standardized scores of the first two FPCs for $K=2,\dots,6$. The maximum ($0.569$) is attained at $K=4$; values for $K=5$ and $K=6$ are close to it.}
\label{fig:silhouette_selection}
\end{figure}

\begin{figure}[htbp]
\centering
\includegraphics[width=0.80\textwidth]{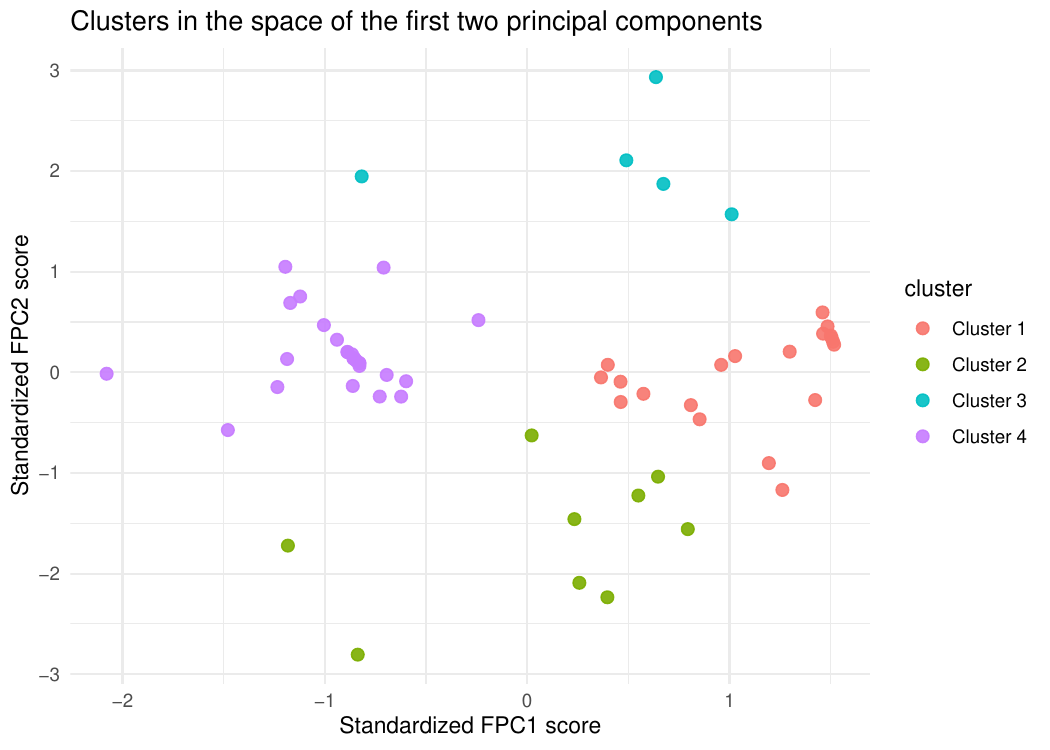}
\caption{Scores of the first two FPCs coloured by cluster. Since $k$-means is applied to these (standardized) scores, the separation of the groups in this plane is a consequence of the clustering and is shown to describe the partition.}
\label{fig:fpca_scores_clusters}
\end{figure}

\begin{figure}[htbp]
\centering
\includegraphics[width=0.75\textwidth]{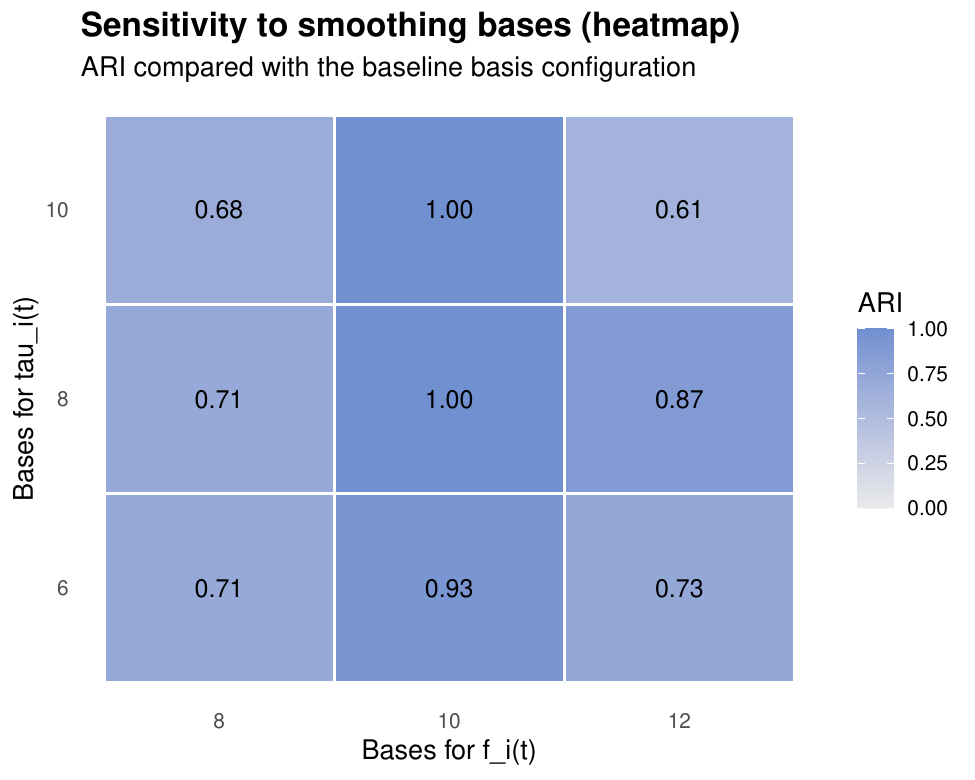}
\caption{Adjusted Rand Index between the reference partition (basis dimension $10$ for $f$ and $8$ for $\tau$) and partitions obtained with alternative basis dimensions for the two smoothing steps.}
\label{fig:robustness_smoothing}
\end{figure}

\end{appendices}

\clearpage

\bibliography{_biblio}

\end{document}